\documentstyle[12pt,aaspp,psfig]{article}
\singlespace
\tighten
\onecolumn
\received{}
\accepted{}
\begin{document}

%%%%%%%%%%%%%%%%%%%%%%%%%%%%   EM's DEFINITIONS  %%%%%%%%%%%%%%%%%%%%%%%%%
\def\sun{\hbox{$\odot$}}
\def\earth{\hbox{$\oplus$}}
\def\lesssim{\mathrel{\hbox{\rlap{\hbox{\lower4pt\hbox{$\sim$}}}\hbox{$<$}}}}
\def\gtrsim{\mathrel{\hbox{\rlap{\hbox{\lower4pt\hbox{$\sim$}}}\hbox{$>$}}}}
\def\slantfrac#1#2{\hbox{$\,^#1\!/_#2$}}
\def\onehalf{\slantfrac{1}{2}}
\def\onethird{\slantfrac{1}{3}}
\def\twothirds{\slantfrac{2}{3}}
\def\onequarter{\slantfrac{1}{4}}
\def \threequarters{\slantfrac{3}{4}}
\def \Point{{\bf $\bullet$ } }
\def \be{\begin{equation}}
\def \ee{\end{equation}}
\def \and{\& }
\def \acknowledgments{\vskip 3ex plus .8ex minus .4ex}
\let \acknowledgements=\acknowledgments			
%%    U N I T S
\def \units{\hspace{1cm}}
\def \erg {~{\rm erg}}
\def \cm{~{\rm cm}}
\def \solarmass{~ M_\odot}
\def \Msun{~ M_\odot}
\def \Lsun{~ L_\odot}
\def \pc{~{\rm pc}}
\def \kpc{~{\rm kpc}}
\def \Mpc{~{\rm Mpc}}
\def \mas{~{\rm mas}}
\def \km{~{\rm km}}
\def \sec{~{\rm s}}
\def \yr{~{\rm yr}}
\def \year{~{\rm yr}}
\def \persecond{~{\rm s}^{-1}}
\def \persec{~{\rm s}^{-1}}
\def \peryear{~{\rm yr}^{-1}}
\def \arcmin{\hbox{$^\prime$}}
\def \arcsec{\hbox{$^{\prime\prime}$}}
\def \micron{\hbox{$\mu$m}}
\def \deg{\hbox{$^\circ$}}
\def \ster{~{\rm ster} }
\def \GHz{~{\rm GHz} }
\def \Kelvin{~\!{\rm K} }
\def \Jy{~{\rm Jy} }
\def \period{\hspace{1cm} . }
\def \coma{\hspace{1cm} , }
\def \bfr{{\bf r}}
\def\excess{\l_{ex}}
\def\meanexcess{\bar{l}_{ex}}
\def\habs{h_{abs}}

\def \los{l.o.s. }
\def \EM{{\rm EM}}
\def \HI{H$_{{\rm I}}$ }
\def \HII{HII }
\def \Hminus{H$^{-}$ }
\def \Halpha{H$\alpha$ }
\def \phisat{\Phi_{\rm sat} }
\def \Htwo{H$_{2}$ }
\def \HtwoO{H$_{2}$O }
\def \water{{\rm H}_{2}{\rm O}}
\def  \Hmol{{\rm H}_{2}}
\def \etal{{\it et~al.~}}
\def \etalb{{\it et}$\!$ {\it al.}$\!$}
\def \eg{{\it e.g.~}}
\def \ie{{\it i.e.~}}
\def \cf{{\it c.f.~}}

\def\Miyo{Miyoshi \etal 1995}
\def\Atha{Athanassoula }

\def \anewpage{}
\def\beforesubsection{\vspace{-0.1truein}}
\def\aftersubsection{\vspace{-0.1truein}}

%%%%%%%%%%%%%%%%%%%%%%%%%%%%% END OF EM's DEFINITION  %%%%%%%%%%%%%%%%
.

\vspace{1.2truein}
 
\title{$\,\,$ Doppler-Shift Asymmetry In High-Velocity Maser \\ $\,$ Emission 
From Shocks In Circumnuclear Disks}

\vspace{0.3in}
\author{Eyal Maoz$^{1,3}$ \, and \, Christopher F. McKee$^{1,2}$ }

\vspace{0.4in}
\affil{$\,\,\,$ $^{1}$Astronomy Department, University of California, 
Berkeley, CA 94720}
\vspace{0.1in}
\affil{$\!\!\!\!\!$ $^{2}$Physics Department, University of California, 
Berkeley, CA 94720}
\vspace{0.1in}
\affil{$\!\!\!\!\!\!\!\!\!\!\!\!\!\!\!\!\!\!\!\!\!\!\!\!\!\!\!\!\!\!\!\!\!\!\!\!\!\!\!\!\!\!\!\!\!\!\!\!\!\!\!\!\!\!\!\!\!\!\!\!\!\!\!\!\!\!\!\!\!\!\!\!\!\!\!\!\!\!\!\!\!\!\!\!\!\!\!\!\!\!\!\!\!\!\!\!\!\!\!\!\!\!\!\!\!\!\!\!\!\!\!\!\!\!\!\!\!\!\!\!\!\!\!\!\!\!\!\!\!\!\!\!\!\!\!\!\!\!\!^3$Miller Fellow}

\vspace{2.4in}
\centerline{$\dagger$ \it Submitted to the Astrophysical Journal }
\vspace{0.4in}

\centerline{$\,\,\,\,\,$ Received:~  \underline{March 4, 1997} ~~ ;~~ Accepted:~ \underline{\hspace{1.97cm}}}

\newpage
%%%%%%%%%%%%%%%%%%%%% ABSTRACT %%%%%%%%%%%%%%%%%%%%%%%%%%%%%%%%%%%%
\begin{abstract}

The rapidly rotating, masing circumnuclear disk in the central sub-parsec 
region of the galaxy NGC 4258 is remarkably circular and Keplerian, yet
a striking asymmetry appears in the maser spectrum: the red-shifted, 
high-velocity sources are much more numerous and significantly more
intense than the blue-shifted ones. A similar strong asymmetry appears 
also in the recently discovered, masing, circumnuclear disks in NGC 1068 
and NGC 4945, thus suggesting it may be a general phenomenon.

We show that the observed Doppler-shift asymmetry can naturally arise due to 
spiral shocks in circumnuclear disks.
We argue that population inversion can largely be quenched in these systems
due to IR photon trapping, and that the high-velocity maser
emission originates within thin slabs of post-shock gas, where the physical 
conditions are conducive to maser action. 
The high-velocity masers with the longest gain paths appear where the 
line-of-sight is tangent to shock fronts.  
Since the spirals have a trailing geometry due to the 
action of differential rotation, the locations of the masers
make the blue-shifted radiation travel through a column of non-inverted 
gas which maintains close velocity coherence with the maser source, where
absorption occurs. 
The resulting asymmetry in the high-velocity maser spectrum, where the
red-shifted emission appears systematically stronger, is independent 
of the existence of a warp in the disk or the azimuthal direction to 
the observer. 

The high velocities of these features reflect the rotational velocities in
the disk, and have nothing to do with the shock speed.
The low-velocity emission arises within a narrow annulus near the inner 
edge of the disk, where direct irradiation by a central source may provide
the energy which ultimately powers the maser. 
In NGC~4258 $-$ the currently most well-defined masing disk $-$ 
the proposed scenario can also account for
the intriguing clustering of the high-velocity 
maser spots in distinct clumps, the restricted spatial distribution of the 
low-velocity sources, and the dip in the maser spectrum 
at the systemic velocity of the disk. In this case we infer a disk mass of
$\sim\!10^4 \solarmass$ and a mass accretion rate of order
$\sim\!7\!\times\!10^{-3} \solarmass \year^{-1}$, which may be 
consistent with an advection-dominated accretion flow.
The model is consistent with the observed Keplerian rotation, and introduces
only negligible corrections to the previously derived black hole mass and
galaxy distance.
Predictions include slow systematic drifts in the velocity and position 
of all the high-velocity features, and the existence of circumnuclear
disks which are delineated only by high-velocity maser emission.

\end{abstract}


\def\totalmass{3.6\!\times\!10^7\,\solarmass }
\def\BH{black hole }
\def \vs{v_s}
\def \trec{t_{\rm rec} }
\def \Ho{Hollenbach }
\def \pitch{\theta_{p}}
\def \pitchdeg{\theta_{p,deg}}
\def \pitchdeg{\theta_{p{\rm ,deg}}}
\def \inclination{\theta_{i}}
\def \beam{\theta_{b}}
\def \Tfront{T_{\nu,{\rm pre}}}
\def \Trec{T_{\nu,{\rm post}}}
\def \los{l.o.s }

\newpage 
%%%%%%%%%%%%%%%%%%%%%%%%%%%%%%%%%%%%%%%%%%%%%%%%%%%%%%%%%
\section{INTRODUCTION}
%%%%%%%%%%%%%%%%%%%%%%%%%%%%%%%%%%%%%%%%%%%%%%%%%%%%%%%%%
VLBI observations by Miyoshi \etal (1995) have revealed that the water maser
emission at the center of the galaxy NGC 4258 arises in a 
rapidly rotating molecular disk ($\simeq\!900 \km\persec$ at $\simeq\!0.2\pc$),
viewed nearly edge-on, surrounding a massive black hole (Maoz 1995a).
The high-velocity maser sources have 
line-of-sight velocities that trace a Keplerian rotation curve to a
remarkably high precision. The fact that the 
red-shifted and blue-shifted high-velocity maser emission
fit into the same Keplerian 
relation indicates that the disk is very nearly
circular. Yet, a striking asymmetry appears 
in the maser spectrum: the red-shifted high-velocity features on the receding
side of the disk are much more numerous and significantly
more intense than the blue-shifted 
ones on the approaching side of the disk, having a total
luminosity ratio of $\simeq23\,$ (Nakai, Inoue \and Miyoshi 1993; \Miyo).
 
Evidence for masing circumnuclear disks has been found also in other
galaxies. The most compelling case is the rotating structure in NGC 1068 
(Gallimore \etal 1996; Greenhill \etal 1996). It is delineated by 
high-velocity \HtwoO maser emission ($\simeq\!300 \km\persec$ at
$\simeq\!0.7\pc$) where, again, the
red-shifted features are much stronger and more numerous than the
blue-shifted ones, being even stronger than the features with the 
systemic velocity of the galaxy (hereafter, the ``systemic'' features). 
A similar situation appears at the center of NGC 4945, where the water 
maser emission consists of systemic features, and features shifted roughly 
symmetrically, in a nearly planar structure, by about 
$\pm(100\hbox{-}150) \km\persec$ (Greenhill, Moran \and Herrnstein 1997). 
A disk-origin of the maser emission is indicated also by the position-velocity
gradient detected in the systemic emission (Greenhill 1996), 
and by the fact that the line connecting the systemic and red-shifted features 
is perpendicular to the axis of the observed ionization cone (Nakai 1989).  
The disk radius is about $\approx\!0.3\pc$, and the binding mass is
$\sim\!10^6 \solarmass$. The red-shifted, high-velocity features are 
stronger than the systemic features, and significantly stronger than
the blue-shifted features (Greenhill \etal 1997).
Evidence for a rotating disk structure 
($\simeq\!100 \km\persec$ at $r\!\simeq\!40 \pc$), delineated by OH 
maser emission, has also been found in the galaxy III Zw 35, located at a 
distance of $\simeq\!100 \Mpc$ (Montgomery \and Cohen 1992). 
Again, the red-shifted maser features are stronger. 
Wilson, Braatz and Henkel (1995) found that the systemic maser emission
from the nucleus of the galaxy NGC 2639 
exhibits a velocity drift similar to that observed in NGC 4258,
suggesting that it too may arise due to centripetal acceleration in a 
rotating disk.  High-velocity maser features, though, 
are yet to be found in NGC 2639.
At present there is no evidence for a masing circumnuclear disk where the
blue-shifted, high-velocity emission is stronger than the red-shifted.
This suggests that the relatively enhanced red-shifted, high-velocity maser 
emission may be a general phenomenon in circumnuclear disks. 
 
High-velocity maser emission, by which we mean emission displaced by
$\gtrsim\!100\km \persec$ from the systemic
velocity of the galaxy, need not necessarily always arise in a rotating disk. 
It may arise in outflows from galactic nuclei, in which case 
the blue-shifted features are likely to appear stronger because the 
radially-approaching side of the flow may amplify a background continuum 
source at the center of expansion, as found in a case of interstellar 
masers (Gwinn 1994). This might be the case in NGC 3079, for example, where 
the maser emission features cover a velocity range which lies completely
below the systemic velocity of the galaxy (Baan \and Haschick 1996).
High-velocity maser emission may arise along 
radio jets, possibly at shock interfaces between the jet and the dense
near-nucleus gas, as found along the jet of NGC 1068 (Gallimore \etal 1996).
If the maser emission originates in the vicinity of a massive black hole, 
whether in a disk, an outflow, or jets, then identifying features as 
``systemic'' or ``satellite'' can be confused by the black hole motion within
the galaxy nucleus: massive black holes may have substantial peculiar 
velocities (\eg Miller \and Smith 1992, and references therein) which would 
shift the maser spectrum in either direction by up to the velocity dispersion
in the galaxy nucleus. 
Therefore, evidence for a disk origin of maser emission must rely on 
both the kinematic and spatial distribution of the maser sources.

Water maser emission has been also detected at the centers of 15 other
galaxies (Braatz, Wilson \and Henkel 1996, and references therein). 
The origin of the maser emission in these galaxies is currently unknown, but
it is reasonable to expect a disk-origin in at least a few of the cases. 
Thus, it is interesting to see whether the red-shifted features in these 
maser spectra tend, in a statistical sense, to be
stronger relative to the blue-shifted ones.
Indeed, the strongest maser features in each spectrum
are red-shifted by $45\hbox{-}200 \km\persec$ relative to the galaxy systemic
velocity in 10 cases (Mrk 1, NGC 1052, NGC 1386, Mrk 1210, IC 2560,
Circinus, ESO 103-G35, TXFS 2226-184, IC 1481, M51), blue-shifted by
$85\hbox{-}150 \km\persec$ in 3 cases (NGC 3079, NGC 5506, NGC 253), and
consistent with the systemic velocity of the galaxy in NGC 2639 and 
NGC 5347. This is compatible with the suggestion that the Doppler-shift 
asymmetry is a general phenomenon in circumnuclear disks, yet,
the above finding should not be regarded as conclusive evidence because (a) 
the origin of the maser emission in these galaxies has not yet been determined,
(b) the asymmetry is not always very pronounced, (c) the relative strength of 
spectral features may vary in time, and (d) the velocities of some of the 
features are not very ``high.'' 

Herrnstein \etal $\!\!$ (1996) proposed that the Doppler-shift asymmetry 
in NGC 4258 may be attributed to the warp in the disk. They suggested that
free-free absorption in the upper atmosphere of the disk attenuates
the maser radiation only from the blue-shifted side of the disk due to 
the orientation of the warp with respect to the line-of-sight. 
Although plausible for an isolated case, such an explanation relies 
on a coincidentally special direction to the observer and is therefore
improbable for a class of circumnuclear disks, particularly
in light of the absence of a clear counter example
(a disk with stronger blue-shifted emission).
Conceivable explanations which rely on maser amplification of non-uniform 
continuum source at the background of the disk, 
or on anisotropic irradiation of the disk by the central source
are improbable for the same reasons.
 
We show that the observed Doppler-shift asymmetry - in which red-shifted
high-velocity maser emission appears stronger than the blue-shifted emission -
can naturally arise due to shocks in circumnuclear disks, independent of the
azimuthal direction to the observer. 
In \S{2} we argue that circumnuclear disks could be largely non-masing
due to IR photon trapping, meaning that maser radiation can be absorbed in 
regions in the disk which maintain 
close velocity coherence with a maser source.
In \S{3} we summarize results of previous investigations which have shown that
the conditions for maser action are favorable behind shock fronts, and argue
that spiral shocks may form in circumnuclear disks for the same reasons 
they do in other astrophysical disk systems.
In \S{4} we describe the geometry of the masing regions behind the spiral
shock fronts, and derive the expected isotropic luminosities of 
red-shifted, high-velocity features.
In \S{5} we show that the trailing geometry of the shock fronts gives rise 
to excess absorption of the blue-shifted, high-velocity maser 
emission, thus producing the Doppler-shift asymmetry. We describe the
geometry of the absorbing regions, and calculate the optical depth for
absorption. 
In \S{6} we discuss in detail 
the remarkably well-defined disk in NGC 4258, where our model can naturally
explain the intriguing spectrum and spatial distribution of the high-velocity 
maser emission, as well as the general properties of the low-velocity emission.
We discuss predictions, and compare our model with that of Neufeld and 
Maloney (1995). 
In \S{7} we discuss the implications of our results to other circumnuclear 
disks in general, and to NGC 1068 in particular. In \S{8} we conclude and
summarize the main points of this investigation.

\def\vlos{v_{los}}

%%%%%%%%%%%%%%%%%%%%%%%%%%%%%%%
\section{A CONDITION FOR INVERTED LEVEL POPULATIONS}
%%%%%%%%%%%%%%%%%%%%%%%%%%%%%%
Maser radiation passing though molecular gas that is at rest relative
to the maser source is either amplified, when the level populations of the 
masing transition are inverted, or absorbed, 
when the level populations are not inverted. 
Inversion can be maintained by a radiative or collisional pumping mechanism,
but the strong extragalactic masers are unlikely to be pumped radiatively
since radiative pumping from continuous-spectrum sources 
requires pumping efficiency of $\sim\!100\%$, which is implausible
(Goldreich \and Kwan 1974, Downes 1983; Genzel 1986; Reid \and Moran 1988). 
The most likely pump for extragalactic $22\GHz$ water masers is collisional. 

Inversion can be achieved only when either the pump or loss rates for the
two levels are different.  In collisional pumping of \HtwoO $\!\!\!$, 
inversion arises due to more spontaneous decays from high-lying states, 
populated by collisions, down to the upper maser level than there are 
down to the lower maser level 
(Hills 1970; De Jong 1973; Strelnitskii 1971; Goldreich \and Kwan 1974). 
However, at densities greater than $\sim \!10^{12}
\cm^{-3}$ collisions re-shuffle the \HtwoO level populations
into a thermal distribution with the kinetic temperature of the gas
(HEM97). Inversion cannot be maintained at such high densities, 
except in cases of inherently non-LTE processes such as collisional pumping by
charged particles that are much hotter than neutrals.
  
The collisional pump in \HtwoO is characterized by the
emission of infrared line photons due to spontaneous decays from 
high-lying states to the upper maser level and to spontaneous decays
from the lower maser level. 
It is essential that these infrared photons not
be re-absorbed by the molecules if inversion is to be maintained.
The IR photons could be removed from the medium either by escaping 
from the masing region, or they may be 
destroyed by being absorbed by cold dust within the masing region.
Thus, the level populations could not be inverted in regions where the minimal
distance to the boundary of the local velocity coherent region is above
some critical value - a scale-length which decreases with increasing density
and water abundance, and increases with increasing abundance of cold
dust. 

The global geometry plays a crucial role 
in determining whether the level populations are locally inverted or not, and
it strongly affects the local maser efficiency. 
Detailed numerical calculations of the equilibrium level populations
in collisional pumping schemes 
(Elitzur, Hollenbach \and McKee 1989, hereafter EHM89;
Hollenbach, Elitzur \and McKee 1997, hereafter HEM97) 
show that the characteristics of \HtwoO maser inversion can be summarized 
quite accurately using the scaling variable $\xi$,
effectively the maser emission measure, defined by 
\be \xi ~\equiv~ {x_{-4}(\water)~ n_{H,9}^2 ~ d_{13} \over \Delta v_5} 
   \label{xi_definition} \coma \ee
where $d_{13}\!\equiv\!d/10^{13} \cm$ is the optical-depth length scale that
controls photon escape, $n_{H,9}\!\equiv\! n_{H}/(10^{9} \cm^{-3})$ 
is the hydrogen
nuclei density, $x_{-4}(\water)\!\equiv\![n(\water)/n_H]/10^{-4}$ is the water
abundance, and $\Delta v_5 \!\equiv\!\Delta v/(10^5 \cm \persec)$ 
is the maser line-width (typically between one a few $\km\persec$).  
These calculations assume static media where
the only velocity gradients present are due to thermal velocities and 
possibly to small-scale chaotic motions (turbulence) in the gas.
Neufeld \and Melnick (1991) performed similar calculations at the presence 
of large-scale velocity gradients (LVG), and found that in the LVG limit (where 
$|dv_z/dz|$ across the masing region
is much larger than $\Delta v/d$), the local equilibrium level
populations are generally a function of the variable $\xi'$, defined as
\be \xi' ~\equiv~ {x_{-4}(\water)~ n_{H,9}^2 \over (dv_z/dz)_{-8}} 
   \label{xitag_definition} \period \ee
This approach enables to treat also cases of media in which the velocity 
gradient is negligible by using the approximate scaling $\xi'\!\simeq\!\xi/9$
(Neufeld \and Melnick 1991).

The pump efficiency, $\eta\!\equiv\!\delta n/n$, which is positive when the
level populations are inverted, falls monotonically with increasing $\xi$ 
in the standard maser parameter space, $0.01\!\lesssim\!\xi\!\lesssim\!10~$, 
and drops abruptly at larger values of $\xi$ as the
maser levels rapidly approach thermalization due to radiative
trapping in the transitions that connect them to other levels. 
Population inversion disappears altogether ($\eta$ becomes negative)
when $\xi$ is above some critical
value $\xi_{crit}$ which depends primarily on the gas temperature
(Neufeld \and Melnick 1990; HEM97). The transition point occurs, for example,
 at $\xi_{crit}\!\simeq\!200$ at $400\Kelvin$, and drops to about $40$ at 
$100\Kelvin~$ (again, notice that $\xi'_{crit}\!\simeq\! \xi_{crit}/9$).
Evidently, in circumnuclear, molecular disks, $\xi$ may exceed 
$\xi_{crit}$ throughout most of the disk volume: molecular densities of $10^9 
\cm^{-3}$ are not outrageously high for these systems
(notice that $n_{H,9}\!=\!2n_{9}[{\rm H}_2]$ for molecular gas); 
the scale-height of the gas distribution in the already discovered masing disks
(\S\S6,7) is a few orders of magnitude larger
than $10^{13}\cm$; and finally, 
the water abundance may by a few times higher than the ISM value
($\sim\!10^{-4}$) due both to the high metallicity in the galactic nucleus, 
as well as 
because copious amounts of water may be produced by X-ray irradiation of the 
disk (Neufeld, Maloney, \and Conger 1994) and by shocks in the disk (\S{3}). 
In such disks, the level populations are generally not inverted, 
except near the inner edge of the disk, which is irradiated directly by X-rays 
from a central nuclear source, and within two, very thin slabs in the 
upper atmosphere of the disk, where the IR photons can escape in 
a short distance vertically.

The effect of cold dust on the radiative transfer would allow larger 
values of $\xi$ before quenching takes place 
(Deguchi 1981; Collison \and Watson 1995). Absorption of IR photons by the dust provides an additional heat sink for collisional pumps, but it requires
that the dust be cooler than the gas. The higher the temperature difference,
the larger is $\xi_{crit}$ above which inversion is quenched. 
We shall assume that the dust and gas are nearly in thermal equilibrium, so 
the presence of dust would not increase $\xi_{crit}$ enough 
to prevent quenching of the inversion in most of the disk volume. 

%%%%%%%%%%%%%%%%%%%%%%%%%%%%%%%
\section{MASER EMISSION FROM SHOCKS IN CIRCUMNUCLEAR DISKS}
%%%%%%%%%%%%%%%%%%%%%%%%%%%%%%%

%%%%%%%%%%%%%%%%%%%%%%%%%%%%%%%
\subsection{Maser Action Behind Shock Fronts}
%%%%%%%%%%%%%%%%%%%%%%%%%%%%%%%

A shock origin of interstellar \HtwoO masers has been suggested by many 
authors (Litvak 1969; Strelnitskii \and Sunyaev 1973; 
Shmeld \etal 1976; Elitzur 1979; Strelinitski 1984; 
Tartar \and Welch 1986; Hollenbach, McKee \and Chernoff 1987;
EHM89; Melnick \etal 1993; Kaufman \and Neufeld 1996; HEM97).
It has been successful in explaining the observed, highly supersonic, radial 
and proper motion velocities of interstellar water maser features, as well
as their high luminosities and small dimensions.  
Detailed modeling revealed that the physical and chemical conditions in 
post-shock regions are favorable for the generation of luminous 22 GHz water 
maser emission by collisional pumping.  
The energy to pump the masers is naturally provided by the dissipation of the 
relative kinetic energies of the shocked and unshocked gas. 
In dissociative, J-type shocks, chemical reactions in the warm, dense 
post-shock gas lead to the production of copious amounts of \HtwoO 
(Elitzur 1979; McKee \and \Ho 1980; EHM89; Neufeld \and \Ho 1994).
High abundances of warm \HtwoO are also produced
behind non-dissociative, MHD (C-type) shocks
($10\!\lesssim\!v_s \!\lesssim\!40\hbox{-}50 \km \persec$) 
which occur in media of low fractional ionization at the presence of
transverse magnetic field (Draine 1980; Draine, Roberge \and Dalgarno 1983;
Chernoff, McKee \and Hollenbach 1983).
The possibility of maser emission from C-type shocks
was first recognized by EHM89, and later confirmed by detailed 
investigations (Melnick \etal 1993; Kaufman \and Neufeld 1996). 
In either type of shock, maser action occurs within a column density of 
$\sim\!10^{22-23}\!\cm^{-2}$ behind the shock front, in 
a slab of thickness $\sim\!10^{13-14}/n_{H,9} \cm$, where the maser emission 
measure is typically $\xi\!\sim\!0.1\hbox{-}10$.
The sheet-like geometry of the shocks allows long gain paths for 
amplification along rays in the shock plane. 
The IR photons produced in the pump cycle easily escape from the masing 
slab through the shock front 
due to the velocity difference between the post-shock and pre-shock gas,  
thus avoiding quenching of the maser
{\it regardless\/} of the dimension of the surrounding gas distribution. 
For all the above reasons, shocks 
provide an ideal location for the generation of strong water masers. 
 
It is important to note that maser action can take place behind shock fronts
only when the pre-shock density does not exceed 
$\sim\!10^{10} \cm^{-3}$, depending exactly on the shock type and strength, 
since otherwise, the post-shock density may reach 
$\approx\!10^{12}\cm^{-3}$ at which H$_2$O masers are
quenched due to collisions (HEM97).
The maximum allowed pre-shock density is higher in the case of C-type 
shocks since, although the total compression in a C-type shock is 
$\sim\!10\hbox{-}20$, the hottest part of the shock occurs where the 
compression in the neutral component is less than a factor of 
2 (Kaufman \and Neufeld 1996).

%%%%%%%%%%%%%%%%%%%%%%%%%%%%%%%
\subsection{On The Presence Of Spiral Shocks}
%%%%%%%%%%%%%%%%%%%%%%%%%%%%%%%

In circular disks, the relative speed between orbits separated by a
distance larger than the disk thickness is supersonic, so 
even small non-axisymmetric disturbances may lead to the formation of
non-axisymmetric shocks, which shear into 
a trailing spiral geometry due to the action of differential rotation.
The existence and importance of spiral shocks were first recognized 
in the context of galactic disks, where shocks were found to develop
naturally, indeed necessarily, in the presence of spiral activity
(Fujimoto 1966; Roberts 1969, 1970; Shu, Milione \and Roberts 1973;
 Woodward 1973; Roberts, Roberts \and Shu 1975). Shocks along the trailing,
slowly rotating spiral waves were suggested to be the mechanism which
triggers gravitational collapse and rapid star formation, explaining 
the observed tendency 
of young star associations and brilliant \HII regions to lie along the 
inner sides of gaseous spiral arms (\eg Roberts 1967; Westerhout 1968).
The formation of spiral shocks and their role in the outward transport
of angular momentum have also been discussed and simulated in the context of
protostellar disks and accretion disks
(\eg Sawada, Matsuda \and Hachisu 1986; Matsuda \etal 1987; Spruit 1987; Spruit
\etal 1987; Morfill, Spruit \and Levy 1993; Chakrabarti \and Wiita 1994). 
 
It is only reasonable to expect that spiral shocks may form in circumnuclear 
disks as well. At present  
there is no observational indication for, or constraint on, the 
prevalence of spiral shocks in these systems. The theory of 
spiral shock generation is quite complex, and current understanding
of it may not be complete. It is clear that shock formation requires
that the disk be irritated, even by weak disturbances of which there is
no shortage at the bottom of a galactic potential well. 
We shall now briefly describe the
two basic mechanisms for spiral shock formation 
which have been thoroughly investigated in the contexts of accretion disks and 
spiral galaxies. We shall point out that circumnuclear disks are not 
much different in the sense of their susceptibility to spiral shock formation. 
The reader who agrees that the existence of spiral shocks 
in circumnuclear disks is indeed plausible may proceed to \S{4}.

Analytical and semi-analytical solutions for shocks in accretion disks
(Spruit 1987) have confirmed earlier conjectures (Lynden-Bell 1974; 
Donner 1979) of the possible generation of self-sustained spiral shocks.
This process requires a weak source of non-axisymmetric disturbances in 
the outer parts of a disk. As a perturbation propagates inward, it is sheared 
by the differential rotation into a trailing spiral pattern and its amplitude
grows. The wave eventually steepens into a shock, with a further increase
in amplitude until further steepening is balanced by dissipation in the
shock (Spruit \etal 1987). From there inward the strength of the shock is
independent of the initial amplitude of the perturbation.
Though shocks in this scenario are set up by some disturbance in the 
outer parts of the disk, they extend all the way in {\it without\/} further 
external forcing.  The theoretically predicted
Mach number of such shocks is never very large, and the angle 
between the shock and the flow is of the order of
$\tan\theta_p \!\sim\!c_s/\Omega r$, which is small in a cool disk 
(Spruit 1987).  The generation and support of such self-sustained,
spiral-shaped shock waves has been observed in simulations
(Sawada, Matsuda \and Hachisu 1986; Sawada \etal 1987).

The second type of mechanism for generating spiral shocks relies on 
the self-gravity of a disk that maintains spiral density waves, 
as in the case of spiral galaxies.
First we should point out the distinction made between 
``grand-design'' spiral galaxies (\eg M51 \and M81), where the
spiral structure is dominated by a few, usually two, major arms, and 
flocculent spirals (\eg Sc galaxies), which can be described
as swirling hotch-potch of bits and pieces of spiral arms
(Goldreich \and Lynden-Bell 1964).
Grand-design spirals are believed to have a well-defined pattern speed,
so material in the disk is orbiting faster than the waves inside corotation,
and more slowly than the waves outside corotation. 
As mentioned earlier in this section, the perturbing gravitational field 
due to spiral density waves was found to be capable of inducing the formation 
of large-scale shocks along the spirals, even if the amplitude of the spiral 
field is only a very small fraction of the axisymmetric field
(Fujimoto 1966; Roberts 1969; Shu \etal 1972; Shu, Milione \and Roberts 1973).
The shocks occur as the orbiting gas slams into the spiral waves which rotate
at a different speed.
The gaseous component of the spiral waves is not confined by 
the inner and outer Lindblad resonances, but can extend throughout the 
entire disk (\eg \Atha 1984).  It is interesting to notice that the pattern
speed in galaxies, where it can be estimated, is typically lower 
than the rotation speed, 
thus placing corotation outside, but not far from, the outer edge of the disk
(Lin 1970; Toomre 1977; Roberts, Roberts \and Shu 1975;
Elmegreen, Elmegreen and Seiden 1989). This could be 
naturally explained if grand-design spirals are triggered by external forcing. 

Flocculent spirals consist of bits and pieces of spiral arms whose
origin lies with the local gravitational instability
(Goldreich \and Lynden-Bell 1965; Julian \and Toomre 1966; Toomre 1981).
The constantly forming and dying spiral waves are triggered by 
imperfections in the disk, such as accidently forming lumps. 
The spiral activity is confined to the range of radii where the disk 
is on the border of stability, namely, where the local value of the Toomre-Q 
parameter is about unity. 
Each spiral wave rotates with the angular speed of its ``irritator,'' so
disk particles inside ``corotation'' overtake the wave due to the differential
rotation, while those orbiting at larger radii are overtaken by the wave. 
Shocks may form along those portions of a wave where the local relative speed 
between the gas and the wave is sufficiently high, 
either inside or outside corotation.
Although the generation of shocks in flocculent spirals 
has not yet been investigated, there is no obvious reason why it would not 
occur as in grand-design spirals (A. Toomre, private communication).

Spiral shocks in circumnuclear disks may arise either due to steepening of 
self-sustained waves, the slamming of gas into spiral density waves (either of
grand-design or flocculent type), or perhaps due to as yet unknown mechanism.
It is clear though
that spiral shock formation depends, in one way or another, on the 
presence of disturbances, either local or at the outer edge of the disk, 
of which there is no shortage at the bottom of a galactic potential well.
Conceivable sources for disturbances
include irregularities in the mass distribution in the
disk and in the mass accretion rate at the disk edge; turbulence due 
to convection (Spruit \etal 1987); magnetic fields; fly-by of 
globular clusters or chunks of giant molecular clouds, and self-gravitating
clumps forming in the disk. The ``outer edge'' of the disk means, in our 
context, the region outside (but not too far outside) the observable disk.  
Obviously, the disk may extend much beyond the masing
part, being undetectable in regions where shocks do not occur or
where the physical conditions are unfavorable for maser action.
In the case of NGC 4258, which we discuss in detail in \S6, the existence
of a massive object orbiting outside the outer edge of the masing disk has 
been recently proposed to explain the origin of the warp in the disk
(Papaloizou, Terquem, \and Lin 1997). If such an object does exist, it might
be the source of the non-axisymmetric perturbation which triggers 
spiral shock formation in this system.

%%%%%%%%%%%%%%%%%%%%%%%%%%%%%%%%%%%%%%%%%%%%%%%%%%%%%%%%%%%%%%%%%%
\section{ MASER EMISSION FROM SPIRAL SHOCKS } 
%%%%%%%%%%%%%%%%%%%%%%%%%%%%%%%%%%%%%%%%%%%%%%%%%%%%%%%%%%%%%%%%%%
\subsection{General Geometry}
%%%%%%%%%%%%%%%%%%%%%%%%%%%%%%%%%%%%%%%%%%%%%%%%%%%%%%%%%%%%%%%%%%
Let us consider a largely non-masing circumnuclear disk that contains spiral 
shocks. The shock fronts can be described, to zeroth-order approximation,
as bent sheets that are as wide
as the disk thickness and perpendicular to the disk plane. 
As elaborated by Roberts (1969), the gas passes through successive 
shocks along its orbit: if the angular speed of the spiral waves is lower 
than that of the disk material, the gas enters each shock front from its 
concave side as pre-shock gas and emerges on its convex side as post-shock gas 
(the other way around if a wave over-runs the gas);
the streamlines nearly repeat themselves through every inter-shock interval, 
being nearly concentric since there must be a small radial transfer of 
mass due to energy dissipation in the shocks.
The shock speed is the flow velocity normal to the spiral in a
frame corotating with the wave. Denoting the angle between the
tangent to the spiral
and the circumferential direction by $\pitch$ (hereafter, the
pitch angle), the shock speed is 
\be v_s ~\simeq ~ r~\!|\omega\!-\!\Omega|\sin\pitch \label{vs} \coma \ee
where $\omega$ is the angular velocity of the unperturbed
flow at radius $r$, and $\Omega$ is the angular velocity of the spiral wave.
  
Assuming the shock speed is high enough above the sound speed and the turbulent
speed in the pre-shock gas to allow efficient IR photon escape through the 
shock front, the level populations of the $5_{23}\!-\!6_{16}$ water maser 
transition are inverted within thin, spiral-shaped slabs of thickness 
$d\!\sim\! 10^{13-14}/n_{H,9} \cm$ behind the shock fronts (\S{3.1}).
The general geometry is illustrated in Figure 1-a, where 
each spiral region represents a masing slab behind a shock 
front, and the horizontal line is the midline $-$ the diameter through the disk
which is perpendicular to the observer's line-of-sight. 
The maser intensity depends on the path length through a velocity 
coherent, inverted region: the mean intensity 
scales exponentially with column density when the maser is unsaturated, 
and linearly when it is saturated. Thus, in the case of a curved slab, 
the most luminous, high-velocity masers
would appear where the \los is tangent to an arc. 
The narrow, dark rectangular in Figure 1-a, 
located where the \los is tangent to one of the masing slabs,
represents such a strong maser. 
Since the maser amplifies only photons produced by spontaneous decays in
the maser itself, there would be no systematic difference between the
intensity of radiation emanating from the two ends of the maser (the 
relativistic beaming effect is negligible at the rotation speeds under
consideration). The maser
photons will appear red-shifted to a distant observer in one direction, and 
blue-shifted to an observer in the diametrically-opposed direction.
 
An immediate consequence of this scenario is that, 
regardless of the azimuthal direction to the observer, the disk should appear
to be delineated by clumps of high-velocity maser spots rather 
than by uniformly or randomly distributed 
spots (maser features may each consist of 
several maser ``spots'' due to micro-turbulence or inhomogeneities
in the emitting region).
It should be made clear that the observed high-velocities of these masers 
reflect the rotational velocities in the disk, and have nothing to do with the
shock speed.

%%%%%%%%%%%%%%%%%%%%%%%%%%%%%%%%%%%%%%%%%%%%
\subsection{Maser Luminosities}
%%%%%%%%%%%%%%%%%%%%%%%%%%%%%%%%%%%%%%%%%%%
\def\Lprod{{\cal L}_{prod} }

The maser luminosity depends on the structure of the shocked region, and thus
on the shock-type. At shock velocities $\lesssim\!50 \km \persec$, shocks
might be either of J-type or C-type, depending upon such poorly known
parameters as the preshock magnetic field and the length scale of the 
ion-neutral coupling. We shall now estimate the expected isotropic luminosity 
of a high-velocity maser feature in either type of shock. We denote the disk
radius at the maser location (where the l.o.s~is tangent to a shock front) by
$r$, and the local thickness of the disk by $h$. 

\vspace{-0.2truein}
%%%%%%%%%%%%%%%%%%%%%%%%%%%%%
\subsubsection{C-type Shocks} 
%%%%%%%%%%%%%%%%%%%%%%%%%%%%%
Kaufman and Neufeld (1996, hereafter KN96) constructed a series of 
models of dense, slow, non-dissociative, C-type (magneto-hydrodynamic) shocks, 
and computed the expected water maser emission. 
The maser luminosity produced per unit area of the shock front is 
$\Lprod \!=\! \epsilon ~\! n_H \mu_{H} v_s^{3}/2$, where $\epsilon$ is 
the efficiency with which shock energy is converted into water
maser emission, and  $\mu_{H}\!=\!2.34\!\times\!10^{-24}~{\rm g}$ is the mass 
of gas per hydrogen nucleus. 
The integration of $\Lprod$ over the surface area of the spiral shock is not
an observable quantity, since measurement of it requires integration over all
lines-of-sight.
However, provided our line-of-sight is typical, this integral is directly 
related to the isotropic luminosity of the maser, $L_{iso}$, which is 
observable ($L_{iso}\!=\!4\pi F_{obs} D^2$, where $F_{obs}$ is the observed
flux from the maser, and $D$ is the galaxy distance).
Since we are interested in the isotropic luminosity of a maser 
at a given radius $r$, we may imagine that the segment of the
spiral shock which emits maser photons in our direction is a part of
a circular masing slab of radius $r$, width $h$, and maser luminosity per
unit area given by $\Lprod$.
Denoting the half beaming angle in the direction normal to the disk 
plane by $\theta_b$, the maser radiation from the ring
is concentrated into a solid angle
$\Delta\Omega\!=\!4\pi\sin\theta_b$ (the maser radiation 
can be detected only by observers that view the disk with inclination larger 
than $90\deg\!-\!\theta_b$). Keeping in mind that the observer's lines-of-sight
are tangent to the circular slab at two points, which would result in two,
equally-bright maser features, the 
isotropic luminosity of a single high-velocity feature at a radius $r$
is $L_{iso} = \!\onehalf~\!(2\pi r h \Lprod) ~\!(4\pi/\Delta \Omega)$.

The fraction $\epsilon_{sat}$ of the mechanical energy in the shock 
that emerges in a water maser transition under the conditions where 
the maser action is saturated has been calculated by KN96, 
assuming cosmic-ray ionization, and a pre-shock magnetic field of 
magnitude $B\!=\!32~\!n_{H,9}^{1/2}~{\rm mG}$ parallel to the shock plane
(such scaling relation corresponds to a constant Alfven speed of 
about $2\km\persec$).  For example, assuming a shock speed of $20 \km
\persec$, molecular hydrogen density of $10^9 \cm^{-3}$, and transverse
magnetic field given by the above scaling relation, KN96 obtained
$\epsilon_{sat}\!\simeq\!10^{-5.5}$, in which case the isotropic luminosity
of a high-velocity maser feature would be  
\be L_{iso} \simeq ~2.7 ~r_{18} ~h_{16} ~n_9(\Hmol)
\left({\epsilon_{sat}\over 10^{-5.5}}\right) \left({v_s \over
20 \km\persec}\right)^3 \left({\theta_b \over 10\deg}\right)^{-1}
\units \Lsun, \label {Ctype-luminosity} \ee
where $r_{18}\!\equiv\!r/10^{18}\cm$, and $h_{16}\!\equiv\!h/10^{16}\cm$.

We should emphasize the limitations of the above estimate for the expected
maser luminosity: 
(a) the fractional ionization may be much higher in the nuclear environment, 
depending on the amount of X-ray irradiation and the vertical column density
of gas in the disk;
(b) assuming other quantities being equal, the energy conversion efficiency
in the case of unsaturated masers, $\epsilon$, is probably different 
than $\epsilon_{sat}$; 
(c) the efficiency $\epsilon_{sat}$ is not an independent variable, 
but depends on the magnetic
field, gas density, and the shock speed; (d) the assumed scaling of the
transverse magnetic field with gas density means that only a part of the
parameter space relevant to circumnuclear conditions has been explored
(it is possible that the Alfven velocity in circumnuclear disks be much 
different than $\simeq\!2\km\persec$).
It is not obvious, at present, how variations in the preshock 
magnetic field, ionization, and grain size distribution will affect the 
emerging maser luminosity, so extending the study of KN96 to include 
other regions of parameter space is most desirable. 
We can only point out that 
the preshock magnetic field and the grain size distribution control the 
thickness and temperature of the shocked layer behind the C-shock front.
Higher magnetic field would result in a shocked region which is thinner and 
hotter. On the other hand, 
the ion-neutral coupling length depends on the grain size distribution, and it
increases if small ($\lesssim\!4\AA$) grains are absent, leading to
a thicker but cooler shocked region. If the fractional ionization is very
high then the shocks would be of J-type rather than C-type. In the case of 
J-type shocks we take a different approach for estimating the expected 
water maser luminosity, as discussed below.

%%%%%%%%%%%%%%%%%%%%%%%%%%%%%
\subsubsection{J-type Shocks} 
%%%%%%%%%%%%%%%%%%%%%%%%%%%%%
In a J-type shock in molecular gas, the temperature just behind the shock
front is $\approx\!5300~\!(v_s/10\km\persec)^2 \Kelvin$ - 
much higher than that in C-type shocks with the same shock speed, and it 
rapidly falls with post-shock column density
(McKee \and Hollenbach 1980; Neufeld \and Hollenbach 1994).
A J-type shock may be either dissociative or non-dissociative. At shock
velocities smaller than $\sim\!15\km\persec$, J-type shocks dissociate only
a small fraction of molecules (Hollenbach \and McKee 1980; Neufeld \and
Hollenbach 1994). As demonstrated by Neufeld and Hollenbach (1994, Fig.~7)
for the case of a pre-shock density of $10^9\cm^{-3}$ and 
$v_s\!=\!10\km\persec$ (non-dissociative shock), 
maser action is collisionally quenched at post-shock 
column densities $\gtrsim\!10^{20}\cm^{-2}$ due to the high compression 
factor. Consequently, the volume of the masing region is much smaller than 
that in a C-type shock, so we expect the maser luminosity to be 
significantly lower than that estimated in \S{4.2.1}.

A dissociative J-type shock, on the other hand,
can produce high intrinsic maser 
luminosity (EHM89). While the temperature profile just behind the shock
front is similar to that in a non-dissociative shock, the heat of H$_{2}$ 
reformation on dust grains provides a substantial energy source which 
maintains a large column of $\sim\!400\Kelvin$ gas, where the chemistry 
drives a considerable fraction of the oxygen not in CO to form \HtwoO$\!\!$.
We now estimate the expected isotropic luminosity of a high-velocity maser
source, assuming a dissociative J-type shock.

Denoting the maximum rate at which maser photons are produced per unit volume
by $\phisat$, the emergent luminosity per unit surface area of the 
shock front is given by 
\be \Lprod \equiv h\nu \int{\phisat ~\! dz} \coma \label{Jtype-1} \ee
where $z$ is the distance behind the shock front, and the integration 
is performed over the post-shock region where the level populations of 
the maser transition are inverted.
The maser photon production rate, $\phisat $, depends on the gas kinetic 
temperature, density, water abundance, and the velocity gradients in the 
masing region. It can be expressed as
\be \phisat ~=~ q ~\!\eta ~\!n_{H}^{2} ~\!x(\water) \coma \label{phi-sat}\ee
where $q$ is the pump rate coefficient, which depends
primarily on $T$ and is almost independent of density. For the water $22\GHz$
transition, $q$ is well fit by (HEM97)
\be q ~\simeq ~ 2.5\!\times\!10^{-11}~ \exp(-644/T) 
\units \cm^3 \persec. \label{q} \ee

We may approximate the integral in  equation (\ref{Jtype-1}) 
by $\phisat(z_p) z_p$, where $z_p$ is the distance behind the shock front
where $z\phisat$ peaks. Multiplying and dividing equation (\ref{Jtype-1}) by
$\Delta v_5$ - the maser linewidth in $\!\!\km \persec$, and substituting
equations (\ref{phi-sat}) and (\ref{xi_definition}), we obtain
\be \Lprod ~\approx~ 10^{27}~\!h\nu ~\! \Delta v_5 ~q(T) ~\! \eta(\xi_p,T)~\! 
\xi_p  \units {\rm erg} \cm^{-2} \persec, \label{Lprod-approximation} \ee
where $\xi_p$ and $T$ are the maser emission measure and 
temperature at a distance $z_p$ behind the shock front.
The pump efficiency $\eta(\xi,T)$ was calculated by HEM97 for $120\!\le\!T\!\le
\!600\Kelvin$: it falls monotonically with increasing $\xi$, and can be fit,
for $\xi\!\lesssim\!10$, by (HEM97)
\be \eta \equiv\delta n/n ~\simeq ~{1\over 20\xi^{1/2} + 0.084 n_{H,9}^{1.3}}
 - {1.05 \over T} \coma \label{pump_efficiency} \ee
which is almost independent of temperature and density. At higher values of
$\xi$, the pump efficiency drops much faster; the deviation from the above
fit begins at a $\xi$ which increases with increasing temperature.
Since the temperature in the H$_{2}$ re-formation plateau is nearly constant 
(EHM89), the maximum of $q\eta\xi$ and that of $\eta\xi$ occurs nearly 
at the same 
$\xi_p$.  Using the numerical results of HEM97, we find that the peak value of 
$\xi\eta(\xi)$ increases monotonically from $0.1$ at $200\Kelvin$
(where $\xi_p\!\simeq\!20.5$ and $\eta[\xi_p]\!\simeq\!5.0\!\times\!10^{-3}$), 
to $0.17$ at $600\Kelvin$ (where $\xi_p\!\simeq\!46$ and
$\eta[\xi_p]\!\simeq\!3.7\!\times\!10^{-3}$). 
Combining these results with 
equations (\ref{Lprod-approximation}) and (\ref{q}), we obtain
\be \Lprod ~\approx~ (0.04|_{200\!\Kelvin}-0.21|_{600\!\Kelvin}) ~ 
\Delta v_5 \units \erg \cm^{-2} \persec, \label{Lprod-res} \ee
where the numerical coefficient corresponds to the temperature range
of $200\hbox{-}600\Kelvin$ explored by HEM97. 
Converting the intrinsic luminosity to 
isotropic luminosity as in \S{4.2.1}, we obtain
\be L_{iso} \simeq ~ (1.9|_{200\!\Kelvin}-9.8|_{600\!\Kelvin}) ~
r_{18} ~h_{16} ~\Delta{v_5}~ \left({\theta_b \over 10\deg}\right)^{-1}
\units \Lsun , \label {Jtype-luminosity} \ee
where the temperature in the H$_2$ re-formation plateau, which determines the
numerical coefficient in equation 
(\ref{Jtype-luminosity}), was estimated by EHM89
to scale with pre-shock gas density and shock speed roughly as 
\be T_{plateau} ~~ \simeq~~ 660 ~ n_9^{2/9}(\Hmol) ~\!
\left({v_s \over 20 \km\persec}\right)^{2/9} 
\left({\Delta v_5 \over 2}\right)^{-2/9} ~\Kelvin. \label{T_plateau}\ee 
The above
derivation assumes a negligible large-scale velocity gradient over the 
masing region, which is a reasonable approximation in the case of J-type shocks
where, unlike in C-type shocks, the velocity change occurs in 
the shock front. 
 
%%%%%%%%%%%%%%%%%%%%%%%%%%%%%%%%%%%%%%%%%%%%%%%%%%%%%%%%%%%%%%%%%%55
\subsection{Structure of the High-Velocity Masers}
%%%%%%%%%%%%%%%%%%%%%%%%%%%%%%%%%%%%%%%%%%%%%%%%%%%%%%%%%%%%%%%%%%55
\def\coher{l_{v,red}}
\def\veldiff{\delta v_{pec}}

The high-velocity maser emission originates in box-shaped regions of thickness 
$d$, height $h$, and length $l_g\!\simeq\!(8{\cal R}d)^{1/2}$, where ${\cal R}$ 
is the local radius of curvature of a spiral (Figure 1-b).
Any monotonically increasing function $\varphi(r)$
describes a spiral with a local pitch angle given by
$\cot\pitch \!=\! r\left|{d\varphi/dr}\right|$, but since 
the global form of $\varphi(r)$ will be clearly irrelevant to our discussion, 
we shall consider for simplicity only 
logarithmic spirals, $\varphi\!=\!\cot\pitch\ln r$, where the pitch angle is 
constant and the local radius of curvature is 
${\cal R}\!=\!r/\cos\pitch$.  Thus, the geometric length of the maser, $l_g$,
defined as the path length through the masing slab, is given by 
\be {l_g\over r}  ~\simeq ~~  10^{-2}~\! \left({d_{13} \over r_{18}}\right)^{1/2} \! (\cos\pitch)^{-1/2} \label{maser_length} \coma  \ee
where $(\cos\pitch)^{1/2}$ is a factor 
of nearly unity, unless the pitch angle is implausibly large.  
The effective length of the maser, which is constrained also by the condition
of velocity coherence, could be smaller than $l_g$, depending on the velocity
field in the disk. Thus, we now address the issue of velocity coherence 
along the \los through a high-velocity maser. 

Let us denote the location of the maser in the disk plane by $(R,y_m)$, 
where $y_m$ is the distance of the maser center from the midline (see Fig.~1b),
and $R$ is the projected distance of the \los from the center of rotation.
Assuming a nearly circular, Keplerian velocity field at the maser vicinity,
the \los velocity along the line-of-sight through the maser,
in the rest frame of the disk,  
is $\vlos \!= \!v_c(R)\left[{1\!+\!(y^2/R^2)}\right]^{-3/4}\!\!$, where 
$v_c(R)$ is the circular velocity at radius $R$. 
Velocity coherence is maintained along the entire geometric length of the maser
if the peculiar \los velocity along the maser does not exceed the maser full 
line-width, $\Delta v$. 
Since the gradient of the \los velocity increases with increasing distance from
the midline, a sufficient condition for velocity coherence is 
$\veldiff\!\lesssim\!\Delta v/2$, where 
$\veldiff\!\equiv\!\left|\vlos(R,y_m)\!~ - ~\! \vlos(R,y_m\!-\!l_g/2)\right|$, 
and $(R,y_m\!-\!l_g/2)$ is the location of the maser end furthest away
from the midline. 
Assuming a logarithmic spiral, which means that $y_m\!=\!R\tan\pitch$, 
expanding $\veldiff$ up to a second order in $\delta y$ (the first order 
vanishes at the midline), and using equation (\ref{maser_length}),
the condition for maintaining close velocity coherence along the entire
geometric length of the maser reads
\be  v_{los,8} \left[{
 {\sin\theta_p \over \sin4\deg} 
\sqrt{d_{13} \over R_{18}} ~~+~~ 0.03 \left({d_{13}\over R_{18}}\right)}\right] 
~~\lesssim~ \Delta v_5 \coma \ee
where $R_{18}\!\equiv\!R/10^{18}\cm$, and
$v_{los,8}\!\equiv\!\vlos/1000\km\persec$.  If the above condition does not 
hold, the effective length of the maser is smaller than $l_g$.
Note that the velocity gradient in the post-shock gas associated with the 
shock itself would not introduce much of a gradient in the \los velocity since
the two directions are perpendicular to each other. 

However, the maser radiation is unlikely to be emitted in a single beam of
vertical half beaming angle $\theta_b\!\simeq\!\tan^{-1}(h/l_g)$ for several
reasons: the thin masing slab $(d\!\ll\!h,l_g)$ is likely to break up 
due to microturbulence into several filaments, each giving rise to a single 
maser spot with typical diameter comparable to the slab thickness;
the shock front may be corrugated (in two dimensions) which would 
fragment the filamentary masers into shorter pieces, depending on the typical 
amplitude and wavelength of the corrugation, and would distort the shape of the
``masing box;''  and the vertically 
stratified structure of the disk may also affect the beaming pattern, 
especially when the disk thickness is much larger then its scale height.
Hence, we leave $\theta_b$ as a free parameter, with the
only obvious constraint that $\theta_b$ must exceed $90\deg\!-\!\theta_{inc}$
for the maser radiation to be detected.
The horizontal extent of each high-velocity maser
clump will be at least as large as the 
uncertainty in the maser positions, which is typically $\simeq\!0.1\mas$, and 
possibly broader than that if the slabs are heavily corrugated 
and thereby effectively thickened (\eg Wardle 1990).

We shall keep adhering
to the working hypothesis that the geometry of the emitting region behind the 
shock fronts is of spiral slabs, as wide as the disk thickness 
and perpendicular to the disk plane, while bearing in mind that it is
only a zeroth-order approximation for a possibly much more detailed and
complex structure.

%%%%%%%%%%%%%%%%%%%%%%%%%%%%%%%%%%%%%%%%%%%%%%%%%%%%%%%%%%%%%%%%%%%
\section{ THE DOPPLER-SHIFT ASYMMETRY}
%%%%%%%%%%%%%%%%%%%%%%%%%%%%%%%%%%%%%%%%%%%%%%%%%%%%%%%%%%%%%%%%%%%
The maser radiation travels through some part of the largely 
non-masing disk along its way to the observer. Since the interaction of maser 
photons with water molecules is a line process, the non-inverted gas 
is transparent to the maser radiation, except in regions in the disk that
maintain close velocity coherence with the maser source, where 
maser photons are absorbed. 

\vspace{-0.13truein}
%%%%%%%%%%%%%%%%%%%%%%%%%%%%%%%%%%%
\subsection{The Absorbing Region}
%%%%%%%%%%%%%%%%%%%%%%%%%%%%%%%%%%%
In a circular disk, $\vlos$ along any \los
is symmetric about the midline: it peaks at $(R,0)$, and drops monotonically 
with increasing $|y|$. This means that in a largely non-inverted disk, maser
radiation originating on one side of the midline at $(R,y_m)$, which goes
through the disk on the other side of the midline, 
will be absorbed at $(R,-y_m)$. The location of such absorbing region is 
shown in figure 1-a as the shaded area above the midline.
Since the spiral-shaped regions of masing gas have a {\it trailing\/}
geometry due to the action of differential rotation, 
the high-velocity masers are always located in 
front of the midline in the receding side of the disk, 
and behind the midline in the 
approaching side of the disk (see Figure 1-a). The Doppler-shift asymmetry 
arises because the blue-shifted maser photons are subject to absorption as 
they pass through the velocity-coherent region on the other side of the 
midline, while red-shifted photons never cross such region since $\vlos$ 
in the disk declines monotonically along their path to the observer. 
 
The red-shifted and blue-shifted beams may also be slightly attenuated just 
outside the maser, where non-inverted gas may maintain close enough velocity 
coherence with the maser source.  Such regions are shown in Figure 1-a 
as the short, shaded areas on both sides of the maser. 
First we shall 
discuss the excess absorption of the blue-shifted maser beam, which
is suggested to account for the Doppler-shift asymmetry in the high-velocity
maser spectrum, and 
then the possible, relatively weaker
absorption of red-shifted radiation just outside the maser.

The first step is to derive the path-length that maser photons travel 
through an absorbing region.
Let us denote the height above the mid-plane of the disk by $z$, where
$-\habs/2\!\le\!z\!\le \habs/2$, $~\habs$ is the disk thickness of the 
absorbing gas, and $z\!=\!\habs/2~$ is the height of the
``upper'' surface of a disk, which 
is directly observable when $\theta_{inc}\!\ne\!90\deg$.
Note that $\habs$ may, in principle, be larger than $h$ - the disk thickness 
of the masing gas behind shock fronts - due to, for example, the vertically 
stratified structure of the disk which may lead to more favorable conditions 
for maser action closer to the disk midplane.
If a disk is viewed exactly edge-on, all {\it detectable\/} rays 
from a high-velocity maser will travel the same path-length through the disk.
However, if the disk is not viewed edge-on, the rays connecting the 
maser and the observer are inclined to the disk plane, so photons leaving 
the maser at different heights in the disk would travel different
path-lengths through the disk: those originating with $z\!=\!\habs/2$
would not pass at all through non-inverted gas, while those leaving the maser
at $z\!=\!-\habs/2$ would travel the largest path-length through the disk.

Denoting the difference in the total path-length of blue-shifted and 
red-shifted maser photons, emanating from the same maser at the same $z$,
through velocity coherent, absorbing
regions by $\excess$, and the location of the maser 
center in the disk plane by $(R,y_m)$ as in \S{4.3}, 
the excess absorption of the blue-shifted radiation occurs within a region
of length 
\be \excess(z) ~\equiv \! \int\limits_{-\infty}^{y_m-(l_g/2)}{\!\!\!\!\Gamma~dy} ~ -  \int\limits_{y_m+(l_g/2)}^{+\infty}{\!\!\!\!\Gamma~dy} 
\label{define-zeta} \hspace{0.8truein} -h/2\!\le\!z\!\le\!h/2 ,
\label{define-excess} \ee
\vspace{-0.09truein}
where 
\be \Gamma(z) ~\equiv~ H\!\left[{{\Delta v\over2} -
 |\vlos(R,y_m)\!-\!\vlos(R,y)|}\right] ~
H\!\left[{{(\habs/2) -z \over|y\!-\!y_m|-l_g/2} - \tan(90\deg\!-\!\theta_{inc})}\right] \coma \label{define-integrand} \ee
and $H(x)$ is the Heaviside step function, which equals unity when
$x\!\!>\!\!0$, and vanishes otherwise. The first term in equation
(\ref{define-integrand}) reflects the condition of velocity coherence, where
$\Delta v$ is the maser line-width.  The second term ensures that the 
integration is carried out only along the path where the 
photons travel through the disk.  As will be 
demonstrated in \S{6.1.3}, it is possible
that blue-shifted rays may pass only through part of the absorbing region 
on the other side of the midline, or even miss it altogether. 

Red-shifted maser radiation may also be absorbed just outside the maser - 
where non-inverted gas may maintain close enough velocity coherence with the 
maser source. Such region is shown in Figure 1-a as the short, shaded area 
below the maser. The length of the absorbing column
is given by the first term in equation
(\ref{define-excess}).  An estimate for the length of the absorbing column
on the red-shifted side of the maser is given by 
$l_{abs,red}\!=\!{\rm max}\left[{ l_{v,red}\!- \!(l_{g}/2) , 0}\right]$, 
where $l_{v,red}$ is determined by the condition of velocity coherence 
$\left|\vlos(R,y_m)\!~ - ~\! \vlos(R,y_m\!-l_{v,red})\right|\!=\!\Delta v/2$, 
and $l_g$ is given by equation (\ref{maser_length}). Clearly, $l_{abs,red}$
vanishes when the non-inverted gas just outside the maser does not maintain 
close enough velocity coherence with the maser source. Approximating 
$l_{v,red}$ by $(\Delta v/2)~\!\!(d\vlos/dy)^{-1}$, assuming 
Keplerian rotation as in \S{4.3}, and substituting $y\!=\!R\tan\pitch$ 
assuming a logarithmic spiral, we obtain
\be l_{abs,red} ~ \simeq ~ {\rm min} \left[{  ~ 
{2 \Delta v~\! R \over 3~\! \vlos \sin(2\pitch)}~
  - ~{l_{g}\over 2} ~  , ~  0 ~}\right] \period \label{lasbred} \ee
This length-scale would typically be much smaller than 
the path-length for absorption of the blue-shifted beam (\eg see \S{6.1.3}). 
We also note that the effective absorption coefficient in this region is
smaller than that along the path of the blue-shifted beam 
since the absorbing region on the other
side of the midline is in velocity coherence with the center of the maser 
line, by definition, while the absorbing region just outside the maser is in 
coherence with the wing of the maser line profile. 
  
The velocity coherent region on the 
other side of the midline may include a small masing section of another spiral 
slab, as illustrated in Figure 1-a. We shall ignore such situation since, if
it happens,  
the length of that masing section will typically be only a few times 
larger than the thickness of the very thin slabs, so it would not reduce much 
the length of the absorbing region. It may only marginally increase the 
effective length of the maser, causing a negligible displacement in the 
location of the maser core.

%%%%%%%%%%%%%%%%%%%%%%%%%%%%%%%%%%%%%%%
\subsection{Optical Depth For Absorption}
%%%%%%%%%%%%%%%%%%%%%%%%%%%%%%%%%%%%%%%

The absorption in the non-inverted gas depends on the level populations of
the maser transition.  Trapping of the IR photons, which quenches the
inversion, would bring the level populations close to their LTE values. But
if the intensity of maser radiation passing through an absorbing region
is very high, it may perturb
the level populations, as the radiation ``tries to invert'' the levels.
This effect is analogous to the process of maser saturation where stimulated
emission starts to diminish the population inversion and so reduces the
magnitude of the (negative) absorption coefficient. Thus, we shall consider 
the cases of saturated and unsaturated absorption of maser radiation separately.

%%%%%%%%%%%%%%%%%%%%%%%%%%%%%%%%%%%%%%%
\vspace{-0.12truein}
\subsubsection{Unsaturated Absorption}
%%%%%%%%%%%%%%%%%%%%%%%%%%%%%%%%%%%%%%%

Following Hollenbach and McKee (1979), 
let $n_l$ and $n_u$ be the populations in the lower and upper levels
of the transition of interest, $g_l$ and $g_u$ the respective
statistical weights, $A_{ul}$ the Einstein coefficient, 
$\lambda_{\micron} \!\equiv\! \lambda/1\micron $ the wavelength, 
$\sigma_5$ the line-of-sight velocity dispersion in units of 
$10^5 \cm \persec$, and $N$ the column density of absorbing molecules. 
The optical depth at line-center is 
\be \tau_{ul} ~=~ 2.2 \!\times\!10^{-19} ~ N ~
\left( { n_l \over n}\right) 
\left({ {g_u \over g_l} { A_{ul} \lambda_{\micron}^3 \over \sigma_5}}\right)
\left({ 1 ~-~ {n_u g_l \over n_l g_u}}\right)  \coma \label{tau_general} \ee
where $(n_{l}/n)$ is the fraction of molecules in the lower state, and
the third term is the correction for stimulated emission.
In the case of
the $6_{16}\!\rightarrow \!5_{23}$ rotational transition in water vapor at
$22.23508 \GHz$, $\lambda\!\simeq\!1.35 \cm$, 
$A\!=\!1.9\!\times\! 10^{-9} \persec$, and the rightmost term in equation 
(\ref{tau_general}) equals $1\!-\!e^{-(\Delta E/KT)} \simeq 1.07/T$ for 
$T\!\gg\!1\Kelvin$. 
Given the energy levels of H$_2$O, kindly provided by David Neufeld, we 
calculated the partition function of water vapor in LTE. It is 
well fit by $Z \simeq 8.79~ T_{2}^{3/2}$
over the range $50\!\lesssim\!T\!\lesssim500 \Kelvin$ to an 
accuracy of $\lesssim\!2\%$, where $T_2 \!\equiv\! T/100\Kelvin$. 
The fraction of molecules in a given state to
all ortho-water molecules is $(2J+1) e^{-(E/KT)}/Z$. Taking an 
ortho-to-para ratio of 3, we obtain $n_{5_{23}}/n_{\water} \!
 \simeq \!0.94 ~\!T_{2}^{-3/2} \!\exp(-6.42/T_{2})$,
where $n_{\water}$ is the density of all water molecules.
The \los velocity dispersion due to thermal broadening alone is 
$\sigma_5 \!\simeq \!0.13(T/A_{\water})^{1/2}\!$, where $A_{\water}$ is the
atomic weight of water, so in order to take the effect of turbulence into 
account we multiply it by the factor ${\rm max}(1,{\cal M})$, where ${\cal M}$ 
is the Mach number of the mean turbulent velocity. 
Denoting the column density of water molecules by $N_{20}(\water)\!\equiv\!
N(\water)/10^{20} \cm^{-2}$, equation (\ref{tau_general}) yields 
\be \tau(5_{23}\hbox{-}6_{16}) ~ \simeq ~ 4.5 ~
N_{20}(\water) ~ \left[{f(T) \over f(100 \Kelvin)}\right]
~ {\rm min}\!\left[{1,{\cal M}^{-1}}\right] \coma \label{tau_specific} \ee
where $f(T) \equiv T^{-3}\!\exp(-642/T)$. The function $f(T)$ increases 
by a factor of $\simeq\!3$ from $100\Kelvin$ to $200\Kelvin$, and 
then slowly drops with further increase in temperature. For $T\!<\!100\Kelvin$
it approaches an exponential decline with decreasing temperature.
We discuss the possible value of ${\cal M}$ for the case of NGC 4258 in 
\S{6.1.3}.

In a stationary medium the line opacity is generally calculated by averaging 
the optical depth, which depends on wavelength, over the line profile. 
However, in the case of absorption on the other side of the midline, the 
optical depth is essentially constant across the line profile, being equal
to its peak value at line-center. The reason for that is the smooth change
of $\vlos$ along the \los which guarantees that 
line-center and off-line-center photons will be 
in coherence with absorbing columns of gas of essentially equal length - 
columns which largely overlap but are 
slightly displaced relative to each other along the line-of-sight.
Thus, the optical-depth at line-center (Eq.~[\ref{tau_specific}])
provides an adequate
approximation for the line opacity.  We note that the above argument also 
means that mild, large-scale deviations from the assumed velocity field in 
the disk would only displace the location of the absorbing column of gas 
on the other side of the midline, but would not affect much the optical
depth.

In our case of maser emission from spiral shocks, we are interested in the
excess absorption of the blue-shifted beams, which depends on the excess 
path-length of the blue-shifted radiation through the absorbing region, 
$\excess$, given by equation (\ref{define-excess}). 
However, as discussed in \S{5.1}, photons leaving the maser at different
heights in the disk travel different path-lengths through the disk, such
that $\excess(z)$ increases with decreasing $z$, where 
$-h/2\!\le\!z\!\le\!h/2$. Instead of integrating the amount of absorbed flux
over $z$, which would require detailed modeling of the stratified structure of
the disk, we shall adopt $\meanexcess\!\equiv\!\excess(z\!=\!h/4)$ as an 
estimate for the 
effective path-length for excess absorption of the blue-shifted beams.
This is not a precise estimate, but it would allow us to demonstrate that
substantial absorption may still occur even for maser photons originating half
way up in the atmosphere of the disk (\S{6.1.3}). 
With this estimate, the optical 
depth for excess absorption of the blue-shifted
radiation, $\tau_{ex}$ is given by equation (\ref{tau_specific})
after substituting $N(\water)\!\simeq\!n^2 ~\!x(\water)~\!\meanexcess$.
 
%%%%%%%%%%%%%%%%%%%%%%%%%%%%%%%%%%%%%%%
\vspace{-0.12truein}
\subsubsection{Saturated Absorption}
%%%%%%%%%%%%%%%%%%%%%%%%%%%%%%%%%%%%%%%
The rate of absorption per unit volume is limited by saturation effects to
$-\phisat$, where $\phisat$ is given by Neufeld and Melnick (1991, Eq.~[16]), 
and is negative when the level populations are not inverted. 
Neufeld and Melnick (1991) demonstrated that the level populations of the 
water maser transition in the LVG limit (\S{2}) are 
a function only of $\xi'$ and $T$ if the relevant nonmasing transitions are 
optically thick. 
They defined a maser rate coefficient $Q\!\equiv\!10^{-14} Q_{-14} \cm^3\persec$
such that the volume emissivity is given by 
\be \phisat ~=~ Q(\xi',T)~\!n(\water) ~\!n ~=~Q_{-14}~\!n_{H,9}^2~\!x_{-4}(\water) 
\units {\rm photons}~\persec \cm^{-3}. \label{phi_satQ} \ee
For example, $Q_{-14}(\xi',100\Kelvin)$ for the water maser transition
becomes negative at $\xi'\!\sim\!5$ and asymptotically approaches a value of 
$-0.0185$ with increasing $\xi'$, and $Q_{-14}(\xi',400\Kelvin)$
becomes negative at $\xi'\!\sim\!24$ and asymptotically approaches a value of 
$-0.324$ (David Neufeld, private communication).

Equation (\ref{phi_satQ}) gives the maximum rate of absorption of water maser
photons per unit volume as a function of gas temperature, density, water
abundance and the amount of IR photon trapping (through the variable $\xi'$).
Thus, in the case of saturated absorption, the maximum fraction of photons 
that can be absorbed is 
\be f_{sat}~=~{-\int{\phisat~\!dl}\over {\cal F}}\coma\label{abs_example}\ee
where ${\cal F}$ is the incident photon flux on the absorbing region, 
and the integration is carried out over the velocity-coherent path-length of 
the beam through non-inverted gas.
However, notice that although the integral in equation (\ref{abs_example}) can
be arbitrarily large, $f_{sat}$ obviously cannot exceed unity since there 
is an upper limit on the true fraction of absorbed flux, $f_{max}$, given by
\be f_{max} ~=~  1 - e^{-\tau} \label{f_abs_max} \period \ee
What happens is that when saturation effects
are important the flux decreases linearly until (if the absorbing slab is
large enough) the flux falls to the level where saturation effects become
unimportant, and from there, the flux decreases exponentially. It is exactly
equivalent to the case of amplification in saturated masers, except in 
reverse.

Let us consider the case where the maser intensity remains high enough
to significantly perturb the level populations along the entire path-length 
of the beam.  The fraction of absorbed photons, $f_{sat}$, is the lowest 
possible since the intensity never falls exponentially within the 
absorbing region.
In order to estimate $f_{sat}$, it is useful to notice that the excess 
absorption of blue-shifted maser beams occurs within a spiral-shaped 
absorbing region located parallel to, but just outside of the convex side 
of the masing spiral slab. The 
thickness of that absorbing spiral region - defined by velocity coherence -
is about $\meanexcess\!\sin\pitch$, assuming the pitch angle is not very large.
Since we are interested in the effect of absorption on the {\it isotropic}
luminosity of a blue-shifted, high-velocity feature, we may imagine that the
absorption of the isotropic luminosity occurs within a spherical shell of 
thickness $\meanexcess\!\sin\pitch$, and a radius slightly larger than the disk 
radius at the maser location, $r$.
Denoting the isotropic luminosity at the absence of absorption by $L_{iso}$, 
which should be comparable to that of red-shifted, high-velocity features, 
we obtain 
\be f_{sat} ~=~ 
{ (4\pi r^2 ~\! \meanexcess\sin\pitch)~\!(-\phisat ~\!h\nu) \over L_{iso}} 
\label{fasb_final} \coma \ee
where $\phisat$ is given by equation (\ref{phi_satQ}).
 
Note that the above expression may exceed $f_{max}$ or even unity, 
which means that a transition from saturated to unsaturated absorption occurs 
somewhere along the path of the beam through the absorbing region.  The
higher $f_{sat}$ above unity, 
the earlier along the path of the beam the transition occurs, and the 
larger is the path-length of exponential attenuation. The fraction of
absorbed photons in such intermediate
cases is clearly high, but deriving the exact value 
requires a detailed modeling of the effect of 
the maser radiation on the level populations as a function of maser
intensity near the transition point from saturated to unsaturated absorption, 
which is beyond the scope of this paper.

%%%%%%%%%%%%%%%%%%%%%%%%%%%%%%%%%%%%%%%%%%%%%%%%%%%%%%%%%%%%%%%%%%
\section{THE MASING DISK IN NGC 4258}
%%%%%%%%%%%%%%%%%%%%%%%%%%%%%%%%%%%%%%%%%%%%%%%%%%%%%%%%%%%%%%%%%%
\subsection {High-Velocity Features}
%%%%%%%%%%%%%%%%%%%%%%%%%%%%%%%%%%%%%%%%%%%%%%%%%%%%%%%%%%%%%%%%%%
\subsubsection{Spatial Distribution and the Keplerian Rotation}
%%%%%%%%%%%%%%%%%%%%%%%%%%%%%%%%%%%%%%%%%%%%%%%%%

The masing circumnuclear disk in NGC 4258 is remarkably well-defined: 
it is delineated over the range of radii $0.16\hbox{-}0.25\pc$ 
by high-velocity water maser sources with \los
velocities ($770\hbox{-}1080 \km\persec$) that trace a Keplerian rotation 
curve around a black hole of $3.6\!\times\!10^7\solarmass$
(Miyoshi \etal 1995; Nakai \etal 1995). The disk is very thin 
($h\!\lesssim\!0.003\pc$), slightly warped (Herrnstein \etal 1996), and 
viewed nearly edge-on ($\theta_{inc}\!\simeq\!83\deg$). 

The high-velocity maser spots are clustered into several distinct 
clumps which are almost equally spaced apart from each other 
(Miyoshi \etal 1995). Nakai \etal (1995) reported on two additional 
high-velocity features in the maser spectrum that were too weak to appear
in the VLBI image. 
Assuming these features also fit in the Keplerian relation, their inferred
locations were found to obey the regularity in the inter-clump
spacing ($\simeq\!0.75\mas$ on both sides of the disk),
making it even more pronounced. 
The possibility that the apparent
clustering of the maser sources is simply due to randomly distributed
gas clumps (or small star forming regions) in the disk
is implausible, not only because of the suspicious regularity in their 
spatial distribution, but also because they all fit a single 
Keplerian rotation curve, meaning that they {\it all} arise near a single
diameter through the disk to within a few degrees, given the accuracy of the
Keplerian fit. 

Maoz (1995b) proposed that the clumps of maser spots are located 
at the intersections of spiral waves with the line that supports the
longest coherent gain paths for amplification in our direction (the midline), 
in a largely masing disk, and are thus spaced apart at the 
characteristic crest-to-crest radial distance between the waves.  
In the present investigation we suggest that the disk in NGC 4258 is 
largely not-masing, and that it contains spiral shocks rather than just 
spiral density waves. In this scenario, high-velocity maser radiation arises
where the l.o.s is tangent to a spiral shock, which naturally
accounts for the high-velocity maser spots being clumped
rather than uniformly or randomly distributed.

We emphasize that the global pattern of spiral shocks in the disk of NGC
4258 cannot be uniquely determined from observations at present
(but see the prediction in \S{6.3})
since it depends on the number of spiral waves, the pitch angle,
the total azimuthal angle subtended by each spiral, and on whether the
spirals are continuous or have gaps. 
It is interesting to notice, though, that assuming the
simplest spiral pattern of
$m$ equally-spaced, winding spirals with a small, constant pitch angle,
$r(\varphi)\!=\!\exp(\varphi\tan\pitch)$,
the radial crest-to-crest distance between
the shocks is $\Delta r\simeq 0.1r(\pitchdeg/m)$, where $\pitchdeg\!\equiv\!
\pitch/1\deg$. This is consistent 
with the observed mean inter-clump separation of $\Delta r/r\!\simeq\!0.1$
if $\pitchdeg\!\simeq\!m$.
Yet, one should bear in mind that these are not the
only possible spiral patterns that can reproduce the spatial distribution of the
maser clumps, and that in any case, the Doppler-shift asymmetry is
independent of the global spiral pattern.

The most remarkable finding about the disk in NGC 4258 is its nearly
Keplerian rotation curve, which indicates
that the high-velocity masers arise along
a diameter through the disk. If the disk was largely masing, that
diameter would be the midline - the major axis of the projected disk
(\eg Elmegreen \and
Morris 1979; Ponomarev, Smith \and Strelnitski 1994). In the shock-origin
scenario proposed in the present investigation, the high-velocity masers are
located
at a distance $|y_m|\!\simeq\!r\sin\pitch$ from the midline, where $r$ is the
disk radius at the position of the maser center.
If the local pitch angle does not vary much throughout the
spiral pattern, all the high-velocity masers will
appear along a diameter through the disk which makes an angle
$\theta_p$ with the midline, and thus trace a (slightly different)
Keplerian rotation curve. This means that the true disk
radius, circular velocity, and
binding mass are higher by a factor $(\cos\pitch)^{-1}$ than the values
derived directly from the projected radius and \los velocities.
This introduces a negligible correction in the case of NGC 4258, where
the typical pitch angle is probably about two degrees (see \S{6.1.2}). 
In any case, the predicted off-midline locations of the masers
are consistent with the observational constraint
that the high-velocity sources must lie within $6\deg$ of the midline,
based on the derived upper limit for their centripetal acceleration
(Greenhill \etal 1995).
The shocks themselves would not introduce substantial deviations from a
Keplerian rotation curve since the velocity jump occurs along the
normal to the shock fronts, which are perpendicular to the line-of-sight at
the positions of the masers.

As pointed out in \S{2}, even in a largely non-inverted disk
the level populations may be inverted within two thin slabs in the upper
atmosphere of the disk, where the IR photons can escape in a short distance
vertically.  The thickness of these slabs would 
be similar to that of the post-shock
masing regions ($\sim\!10^{13-14}/n_{H,9} \cm$), which is much smaller than the
disk thickness.   Although amplification along rays within these slabs may
produce strong masers, these masers are unlikely to be detected 
because their length, defined by velocity coherence, is 
so much larger than the thickness of these slabs, that the vertical
beaming angle would be very small.  Such masers could be detected 
only in regions in the disk where the locally tangent plane is viewed
edge-on. It is improbable that the observed high-velocity maser 
emission in NGC 4258 arises in these slabs since it would require wiggles or
wrinkles in 
the disk that happen to bring the extremely narrow beams into sight
only along a single diameter through the disk (as indicated by the Keplerian
rotation), on both side of the disk, and only along that line. 

%%%%%%%%%%%%%%%%%%%%%%%%%%%%%%%%%%%%%%%%%%%%%%%%%
\subsubsection{Shock Speed and Luminosity of the Red-Shifted Features}
%\vspace{-0.02truein}
%%%%%%%%%%%%%%%%%%%%%%%%%%%%%%%%%%%%%%%%%%%%%%%%%
The shock speed must be higher than the sound speed in the gas, $c_s$.
Given the condition of hydrostatic equilibrium, $H/r\simeq c_s/v_c$, 
where $H\!\lesssim\!h/2$ is the disk scale-height, and
the observed upper limit to the thickness of the masing disk, 
$h\!\lesssim\!10^{16}\cm~$ (Miyoshi \etal 1995), the upper limit to the
sound speed at the mean radius of the disk ($r\!=\!0.2\pc$, $v_c\!=\!
900\km\persec$) is $c_s\!\lesssim\!15\km\persec$.  It should be emphasized
that magnetic pressure may play an important role in the vertical support of
the disk, in which case the sound speed could be much lower than 
$15\km\persec$.  This may well be the situation in the disk of NGC~4258 
where the typical maser line-width is only a few $\km\persec$. 
Thus, a reasonable lower limit to the required 
shock speed in the disk is $\approx\! 10\km\persec$. 
The shocks might be either of J-type or C-type, depending upon such
poorly known parameters as the preshock magnetic field and the fractional
ionization. In the case of J-type shocks, significant maser luminosity is
produced only if a large fraction of the molecules are dissociated 
(\S{4.2.2}), which requires $v_s\!\gtrsim\! 20\km\persec$. 
Thus, we shall assume that the shock speed in the masing disk of
NGC 4258 is at least $20\km\persec$, in which case the conditions for maser 
action are favorable behind shock fronts in either type of shock (\S{4}).

The shock speed and pitch angle of a spiral are locally related through the
pattern speed and rotational velocity in the disk (Eq.~[\ref{vs}]). 
Assuming the angular speed of the spiral waves is of order
half the disk rotation speed, as expected in cases where the spirals
originate in disturbances at the outer edge of the disk (\S{3}), we obtain 
\be \pitch ~\approx~ 2.5\deg \left({v_s \over 20 \km\persec}\right)
\left({ v_c \over 900 \km\persec}\right)^{-1} \period\label{shock-speed} \ee
Thus, given the uncertainties in the angular speed of the spiral waves and 
in the shock speed, we expect in NGC~4258 a typical pitch angle of 
order two degrees, assuming a shock speed of about $20\km\persec$. 

Substituting $h\!\lesssim\!10^{16}\cm$ and 
$\theta_b\!\gtrsim\!90\!-\!\theta_{inc}\!=\!7\deg$ (Miyoshi \etal 1995)
in equations (\ref{Ctype-luminosity}) and (\ref{Jtype-luminosity}), the 
expected isotropic luminosity of a red-shifted, high-velocity feature at
the mean radius of the disk is, in the case of C-type shocks, 
\be L_{iso} \lesssim ~2.3 ~n_9(\Hmol)
\left({\epsilon_{sat}\over 10^{-5.5}}\right) \left({v_s \over
20 \km\persec}\right)^3 \units \Lsun, \label{Ctype-lum4258} \ee
and in the case of slow, dissociative, J-type shocks, 
\be L_{iso} \lesssim ~ 8.4|_{600\!\Kelvin} ~
~\Delta{v_5} \units \Lsun, \label{Jtype-lum4258}  \ee
where $\Delta v_5$ is typically between one and a few. The
numerical coefficient in equation (\ref{Jtype-lum4258}) is for a temperature of 
$600\Kelvin$ in the H$_2$ reformation plateau (the approximate dependence of
$T_{plateau}$ on gas density and shock speed is given by Eq.~[\ref{T_plateau}]).
The observed, mean isotropic luminosity of the red-shifted, high-velocity 
features in NGC 4258 is about $4\Lsun$ (Nakai \etal 1995), 
which can be easily reproduced in the case of dissociative, J-type shock
(Eq.~[\ref{Jtype-lum4258}]), and is marginally consistent with C-type 
shocks, given the uncertainties in the magnetic field, fractional ionization, 
estimated energy conversion efficiency, and shock speed. 

%%%%%%%%%%%%%%%%%%%%%%%%%%%%%%%%%%%%%%%%%%%%%%%%%
\subsubsection{Absorption of the Blue-Shifted Radiation}
\vspace{-0.02truein}
%%%%%%%%%%%%%%%%%%%%%%%%%%%%%%%%%%%%%%%%%%%%%%%%%
The asymmetry in the high-velocity maser spectrum of NGC 4258
is striking: the blue-shifted
features are systematically weaker than the red-shifted ones by about an
order of magnitude (Nakai, Inoue \and Miyoshi 1993; Miyoshi \etal 1995).
We now check whether such substantial absorption is consistent with our model.

First, we calculate numerically the velocity coherent path-length for excess 
absorption of the blue-shifted maser radiation, $\excess$, 
at the mean radius of the masing disk (Eq.~{\ref{define-excess}]).  
The result is presented in Figure 2 as a function of pitch angle for three
different choices of $z$ - the height in the disk atmosphere where the 
photons emanate from the maser. This calculation
assumes a disk thickness of $h_{abs}\!=\!0.003\pc$ for the absorbing
gas, $d\!=\!10^{13}\cm$ and consequently
$l_g\!\simeq\!8\!\times\!10^{15}\cm~$ (Eq.~[\ref{maser_length}]), and maser
line-width of $2\km\persec$ ($\Delta v_5$ is typically between one and a few).
The three curves in Figure 3 show that maser photons emerging from
the maser closer to the upper surface of 
the disk pass through a smaller column of absorbing gas.  The solid line, which
corresponds to maser photons emanating from the maser at the midplane of the
disk, gives the mean path-length for excess absorption of blue-shifted photons.

The sharp peak in these curves corresponds to the situation where the maser 
is so close to the midline that the velocity coherent regions on both side 
of the midline touch and connect into one. The steep fall-off at larger 
pitch angles is because the maser is too far from the absorbing region on 
the other side of the midline, so the l.o.s.~through the maser, which is 
inclined to the disk plane by $\simeq\!7\deg$, does not pass through the 
absorbing region.  
The effect of larger maser line-width or larger disk thickness 
would be to broaden the range of $\pitch$ where $\excess(z)$ is substantial.

The amount of absorption strongly depends on the gas temperature, which is 
quite uncertain.  The fact that the disk is slightly warped and hence 
obliquely irradiated by the central source provides a lower limit to the 
mean gas temperature in the disk. 
The X-ray luminosity of the nucleus in the $2\hbox{-}10$ keV range is 
$L_x\! \simeq\! 4\!\times\! 10^{40} \erg \persec$ (Makishima \etal 1994), and
the warping angle of the disk is 
$\theta_{w}\!\lesssim \!20\deg$ (Herrnstein \etal 1996).
For direct attenuated radiation, about one-third of the incident radiation
is absorbed (Cunningham 1976), so the black-body temperature of the disk
due to the X-ray irradiation is given by 
$\sigma T_{irr}^4\! \approx\! \onethird L \sin\theta_{w} / (4\pi r^2)$,
which yields $T_{irr} = 60\Kelvin~(\sin\theta_{w}/ \sin 20\deg)^{1/4}$ at
the mean radius of the disk.
The mean gas temperature is likely to be somewhat higher than that because 
there are other possible heat sources such as direct irradiation by 
Compton-heated corona, and viscous energy dissipation in the disk.  
Moreover, if the absorbing column is not too distant from the shock
front, which happens when the pitch angle is small, the gas 
in the absorbing region may not have completely cooled down to its
pre-shock temperature (notice that in Fig.~1a, 
$\pitch\!=\! 20\deg$ for a clearer illustration, while we expect
$\pitch\!\approx\!2\deg$ in the case of NGC 4258).

Replacing $N({\rm H}_2{\rm O})$ in equation
(\ref{tau_specific}) by $n({\rm H}_2) x(\water) \meanexcess$, and substituting
$\meanexcess\!=\!10^{16}\cm$ (defined in \S{5.2.1}), the optical depth 
for excess absorption of the blue-shifted radiation in the unsaturated
case (\S{5.2.1}) is
\be \tau_{ex} ~ \approx ~ 32~ n_{9}({\rm H}_2)~x_{-4}(\water) 
~\left[{f(T) \over f(100 \Kelvin)}\right] ~ {\rm min}\!
\left[{1,{\cal M}^{-1}}\right] \label{tau-in-4258} \coma \ee
where the mean Mach number of the turbulent velocities, ${\cal M}$, is 
probably of the order of a few, and is certainly limited by the condition 
of hydrostatic equilibrium to 
${\cal M} < (v_c h/2r) ~\!/~\! (0.04 T^{1/2}\km\persec) 
\simeq 18 ~\!(T/100\Kelvin)^{-1/2}$. Again, notice that if magnetic pressure
plays an important role in the vertical support of the disk then ${\cal M}$ 
could be much lower than the above limit. We also note that 
the gas density in the absorbing region could be higher than
the pre-shock (or mean) density in the disk since, when the pitch angle is 
small, the absorbing region is not very far behind the shock front
(relative to the inter-shock spacing). 
Thus, we conclude from equation (\ref{tau-in-4258}) that the optical depth 
for unsaturated absorption in the disk of NGC 4258 can be
easily high enough to attenuate the intensity of the blue-shifted beams
by roughly an order of magnitude, as observed. We now examine whether 
substantial attenuation may occur also in the case of saturated absorption
(\S{5.2.2}).

              %%%%%%%   SATURATED   ABSORPTION  %%%%

The rate of absorption per unit volume, in the limit where saturation effects
are important, depends on the amount of IR photon trapping 
through the variable $\xi$, defined in equation (\ref{xi_definition}).
Assuming pre-shock gas density of $n({\rm H}_2)\!\sim\!10^9\cm^{-3}$, 
disk thickness of $h\!\sim\!10^{16}\cm$, and expected water abundance in
the galactic nucleus of one to a 
few times $10^{-4}$, we obtain $\xi\!\sim\!10^3~$ throughout most of 
the disk volume (notice that $n_{H,9}\!=\!2n_9[{\rm H}_2]$). 
Assuming that the velocity gradient in the absorbing region is small, which
is an adequate assumption in the case of J-type shocks, we can use the 
results derived in the LVG limit (David Neufeld, private communication), where 
$\xi'\!\simeq\!\xi/9\!\sim\!100~$.
Substituting $Q_{-14}(100\Kelvin)\!\simeq
\!0.02$  and $Q_{-14}(400\Kelvin)\!\simeq\!0.02$ for $\xi'\!\sim\!100$
(David Neufeld, private communication), equation (\ref{phi_satQ}) gives 
\be \phisat ~\simeq~ (0.08|_{100\Kelvin}\!-\!0.8|_{400\Kelvin})~\!n_9^2(\water)
~\!x_{-4}(\water) \units {\rm photons}~\persec \cm^{-3}. \label{phi_satQ4258} 
\ee
Substituting $\meanexcess\!=\!10^{16}\cm$ in equation 
(\ref{fasb_final}) we obtain 
\be f_{sat} ~\simeq~ (1.1|_{100\Kelvin}\!-\!11|_{400\Kelvin})~
 \left({\meanexcess\over 10^{16}\cm}\right)~\!
\left({\pitch\over 2\deg}\right)~\!
 ~\! \left({L_{iso} \over 4\Lsun}\right)^{-1} \coma \ee
which is above unity; as discussed in \S{5.2.2}, this indicates that the 
absorption has become unsaturated (note that we have assumed 
that the typical isotropic luminosity of the blue-shifted, 
high-velocity features would have been comparable to that observed for 
the red-shifted ones, $\simeq\!4\Lsun$, at the absence of absorption.) 
Therefore, we conclude that our model is consistent with the intensity of
the blue-shifted, high-velocity maser beams in NGC 4258 being significantly
lower than that of the red-shifted ones.

We note that the path length for possible absorption of the red-shifted
beams in NGC 4258, given by equation (\ref{lasbred}) and evaluated at the 
mean radius of the masing disk, is $\lesssim\!3\times\!
10^{15} \cm$, which is about an order of magnitude smaller than the path length
for absorption of the blue-shifted beams. Combined with the fact that this
relatively small absorbing column is in velocity coherence with the wing of
the maser line rather than with the line-center (\S{5.1}), we conclude that
any possible attenuation of the red-shifted, high-velocity beams in NGC 4258
is negligible. 

%%%%%%%%%%%%%%%%%%%%%%%%%%%%%%%%%%%%%%
\subsection{The Low-Velocity Features}
%%%%%%%%%%%%%%%%%%%%%%%%%%%%%%%%%%%%%%
\vspace{-0.12truein}

The strongest features in the maser spectrum of NGC 4258 
arise near the \los to the center of rotation, and have low velocities with
respect to the systemic velocity of the disk (Miyoshi \etal 1995). 
If the gas in the disk is largely not inverted, as we argue in the present 
investigation, what is the origin of the low-velocity maser emission?  
It is unlikely to arise in the masing spiral slabs because these are very thin
($\sim\!10^{13-14}/n_{H,9} \cm$),
and are nearly perpendicular to the lines-of-sight under consideration.
Before addressing that question, we point out two intriguing 
details in the spatial distribution and spectrum of the low-velocity 
maser emission.

Close velocity coherence is maintained along the entire length of the 
minor axis of the projected disk. Yet, all the observed low-velocity maser 
sources are confined to a narrow annulus of radius $0.13 \pc$, 
inner to the region where high-velocity sources are detected 
($0.16\hbox{-}0.25 \pc$). The width of that annulus is 
$\delta r\!\lesssim\!0.013\pc$, 
as indicated by the dispersion in the centripetal acceleration of the 
maser spots (Greenhill \etal 1995). 
It is difficult to understand the restricted spatial distribution of 
the low-velocity masers if the entire disk is masing. Miyoshi \etal (1995) 
suggested that it may result from the amplification of a continuum source 
of finite extent at the center of rotation, which appears - due to the disk 
inclination - only at the background of the inner part of the disk 
(see illustration in Fig.$\!$ 20 of Nakai \etal 1995). 
Although the low-velocity masers may well amplify a central continuum 
source, the suggested explanation for their restricted distribution is
somewhat improbable since it requires 
precise fine tuning: the effective size of the background continuum 
source must coincide with the projected distance of the inner 
edge of the disk to the center, 
to within $\simeq 0.013\cot\theta_{inc} = 0.0016\pc$. 
Also, if the entire disk is masing, it is unclear why no high-velocity 
features were found at $0.13\!\le\!r\!<\!0.16\pc$ on either side of the disk. 

In a largely masing disk, the low-velocity maser emission should generally
peak at the systemic velocity of the disk, and decline with increasing 
relative velocity.
The reason for that is twofold:  the gradient of the \los velocity vanishes 
along the minor axis of the projected disk, or in the case of an edge-on disk, 
along the \los to the center, and it increases with increasing impact parameter
$b$.  Thus, the velocity coherent gain path for amplification in our 
direction drops with increasing $b$, which should result in a diminishing
$F_\nu$ with increasing $|\vlos|$.
In addition, the surface brightness profile of a background continuum source 
is likely to be centrally concentrated, thus making 
the expected peak at $b\!=\!0$ even more pronounced. However, 
in NGC 4258, instead of a peak there is a striking {\it dip\/} in the 
maser spectrum at the systemic velocity of the disk, which splits the 
low-velocity complex into two distinct components.  The flux density at the 
systemic velocity is weaker even than that of the red-shifted, 
high-velocity features (\eg Greenhill \etal 1995). 
The dip is unlikely to simply be just a fluctuation because of its 
prominence, its location
at the systemic velocity of the disk, and because it persists in spite of 
the variability of the entire low-velocity complex (Nakai \etal 1995).  
Watson and Wallin (1994) suggested that absorption in the disk
may play an important role in splitting the low-velocity maser emission 
complex in NGC 4258 at the systemic velocity.

We suggest that the low-velocity maser emission in NGC 4258 arises in
a ring of masing gas at the inner edge of the disk, where the direct
X-ray irradiation by the central source may provide the energy that 
ultimately powers the maser (\eg Neufeld, Maloney, \and Conger 1994).  
We now argue that the existence of a narrow, 
masing annulus at the inner edge of the largely non-masing disk can naturally 
explain the intriguing findings discussed above. 

Low-velocity maser radiation from the inner edge of the disk is subject 
to absorption as it passes through the largely non-masing disk.
The length of the absorbing column of gas in the foreground of the low-velocity
masers is maximal at the projected minor axis of the disk,
and drops rapidly with increasing impact parameter $b$. 
The absorbing column extends from the outer boundary of the 
masing annulus, along the l.o.s in our direction, out the point where 
$v_{los}$ in the disk deviates from that of the maser source by the maser 
line-width ($\Delta v$). Denoting the radius and circular 
velocity of the masing annulus by $r'$ and $v_{c}'$, respectively, 
it is simple geometry to derive the length of the velocity-coherent,
absorbing column $\l'(b)$, 
or $\l'(\vlos)$. In the case of an edge-on disk with a Keplerian rotation
curve we obtain $\l'(\vlos)\!\simeq\!
(r'/4)(\Delta v/|\vlos|)$ for $\Delta v\!\lesssim\!|\vlos|\!\lesssim\!v_c'/2$
(notice that it is independent of the circular velocity at the maser). 
When $|\vlos|\!<\!\Delta v$, $~l'(\vlos\!\!\rightarrow\!0)\longrightarrow
 (r_{max}\!- \!\!r')$, where $r_{max}$ is the maximum radius of the disk. 
However, as discussed in \S{5.1}, in a disk that is not viewed exactly 
edge-on there is an
upper limit to the path length of any \los through the disk, which is 
determined by the disk thickness $\habs$ and inclination $\theta_{inc}$. Thus,
maser photons emanating in our direction
from a low-velocity maser spot at a height $z$ in
the disk atmosphere (see definition in \S{5.1}) pass through an absorbing 
column of length 
\be  \l'(\vlos) ~ \simeq~ min \left[{ ~{r' \Delta v\over 4 |\vlos|} ~~,~~ 
 \left({{\habs\over 2}-z}\right)\tan\theta_{inc} ~~,~~
r_{max}\!\!-\!r'~  }\right] \label{lowv-zeta} \period \ee 
Figure 3 presents $\l'(\vlos)$ in the case of NGC 4258, 
where $r'\!=\!0.13\pc$, assuming $\Delta v\!=\!2 \km\persec$, 
$\habs\!=\!0.003\pc$, and $\theta_{inc}\!\!=\!83\deg$.
The absorption of low-velocity maser emission is maximal around the 
systemic velocity of the disk, and drops rapidly with increasing relative 
velocity. We suggest that this may be the reason for the prominent dip in the 
low-velocity maser spectrum near the systemic velocity of the disk.

The reason for the concentration of the low-velocity sources within a narrow
annulus at the inner edge of the disk is now also obvious. 
We have to explain, though, the absence of high-velocity emission at
the intersections of the masing annulus and the midline.  The reason for that
is two fold:  the velocity coherent path length in the masing annulus
for amplification in the tangential direction is $l_{tan}\!\simeq\!0.004\pc$,
derived from the condition 
$|\vlos(r',0) - \vlos(r',l_{tan})|\!=\!\Delta v/2$ as in
\S{4.3}, while the path length for amplification in the radial 
direction, which is the width of the annulus, is $l_{rad}\!=\!
\delta r\lesssim\!0.013\pc$, 
as estimated from the dispersion in the centripetal acceleration of the 
maser sources.  Thus, $l_{rad}$ may well be larger than $l_{tan}$, in which
case the velocity coherent regions within the masing annulus are elongated 
in the {\it radial\/} directions. The radially propagating beams will then
win in the competition for available pump photons over beams in the locally
perpendicular direction (the direction tangent to the annulus).
Moreover, the annulus of inverted gas is clearly illuminated by 
a central source of continuum radiation - 
presumably the bisymmetric nuclear jet that has been recently
detected on milli-arcsecond scale in radio continuum (Herrnstein \etal 1997), 
and on much larger scales in 
H$\alpha$ (Ford \etal 1986; Cecil, Wilson \and Tully 1992), 
radio continuum (van der Kruit, Oort \& Mathewson 1972), and in X-ray 
(Cecil, Wilson \& De Pree 1995). Further evidence for amplification 
of a central source comes from 
the absence of low-velocity masers along the minor axis of the projected disk
on the {\it rear} side of the disk, in contrast to the
strong low-velocity emission observed along the same axis
on the front side of the disk, although the two regions equally maintain
velocity coherence. 
The central continuum source provides ample seed photons to be amplified 
in the {\it radial\/} directions, while beams in the locally
perpendicular directions only amplify photons generated due to 
spontaneous decays in the masing region. For these two reasons, 
the maser radiation within the masing annulus at the inner edge of the disk
may be strongly beamed in the radial directions, so it would not be
detectable where lines-of sight are tangent to the ring. 

\newpage
%%%%%%%%%%%%%%%%%%%%%%%
\subsection{Prediction}
%%%%%%%%%%%%%%%%%%%%%%%
\vspace{-0.12truein}
 
The shock-origin scenario has a unique observational consequence. 
The high-velocity masers arise where the \los is tangent to the 
spirals, so the slow rotation of the waves must result in a continuous,
outward drift in the positions of all the high-velocity maser clumps. Assuming 
a rotating logarithmic spiral, 
$r(\varphi,t)\!=\!\bar{r}\exp[\tan\pitch(\varphi\!+\!\Omega t)]$, 
where $\Omega$ 
is the angular velocity of the waves at the mean radius of the disk $\bar{r}$, 
the proper motion of the high-velocity masers is given by 
$\Omega \tan\pitch \bar{r}/D$, where $D$ is the galaxy distance. Assuming 
that $\Omega$ is about half the angular velocity of the disk, we predict 
a systematic proper motion of $\approx\!0.25~\! (\pitch/1\deg) ~\mu{\rm as} 
\peryear$. The slow increase in the distance of the high-velocity clumps 
from the center of rotation must be accompanied by a slow decrease in their
 \los velocities at a rate $dv/dt \!=\! \onehalf v\Omega 
\tan\pitch$, which yields $\simeq 0.02 (\pitch/1\deg) \km\persec\peryear$. 
The outermost maser clump will eventually disappear, and a new high-velocity 
clump will appear at the inner edge of the disk, sometime within the next 
$\approx\!3000 (\pitch/1\deg)$ years, 
given the observed mean separation between the clumps ($\simeq\!0.75$ mas). 
At present, the 
predicted drift rates in position and velocity are too small to 
be detected for individual maser spots. However, since the 
drifts are systematic, it may perhaps be possible to test their existence in 
a statistical analysis of the ensemble of high-velocity maser spots. 

%\newpage
%%%%%%%%%%%%%%%%%%%%%%%%%%%%%%%%%%%%%%
\subsection{Comparison with the Model of Neufeld and Maloney}
%%%%%%%%%%%%%%%%%%%%%%%%%%%%%%%%%%%%%%
\vspace{-0.12truein}

Neufeld and Maloney (1995; hereafter NM95) proposed that the location and 
extent of the masing region in NGC 4258 is determined by the X-ray irradiation 
of the warped disk by the central source. According to their model, the 
circumnuclear disk is completely atomic outside some critical radius,
where the ionization parameter exceeds a critical value. Inside that radius,
the disk becomes increasingly molecular, where warm molecular gas near the disk 
midplane is ``sandwiched'' by hot atomic gas in the disk atmosphere. 
The level populations of the water maser transition are inverted throughout the
warm molecular gas, so the high-velocity masers arise along the midline of
the disk which supports long coherent paths for amplification in our direction.
Assuming that the observed outer radius of the masing disk ($0.25\pc$) 
represents the critical radius beyond which molecules may not exist, NM95 
predict molecular hydrogen density of $3\!\times\!10^7 \cm^{-3}$ in the masing
disk, and total disk mass of $\sim\!100 \Msun$.  
The inner edge of the masing region is suggested to represent the
point at which the warp flattens out and the disk can no longer be heated 
by obliquely-incident X-rays, so the temperature of the molecular gas
at small radii is too low to produce significant water maser emission.

In the model presented in the present paper, the molecular density in the 
masing disk is $\sim\!10^9 \cm^{-3}$, and the disk mass is consequently 
$\sim\!10^4 \Msun$. The gas is molecular throughout the region under 
consideration due to the larger shielding column density of hydrogen 
nuclei, except perhaps high in the disk atmosphere. 
The level populations of the water maser transition are largely not inverted
due to IR photon trapping.  High-velocity maser emission arises behind
spiral shocks in the disk, where the conditions are conducive for maser action.
The observed high-velocity maser features originate where the line-of-sight is
tangent to spiral waves, where the path-length through velocity coherent,
inverted gas is the largest. The high-velocity masers appear nearly along
a diameter through the disk which, in the case of NGC 4258, makes an angle
of about $2\deg$ with the midline (it does not introduce a noticeable deviation
from Keplerian rotation).  

The NM95 model accounts for the observed extent of the masing 
disk, while in the present 
model, high-velocity features are detectable in the region
in the disk which maintain spiral activity, where the strength of 
spiral shocks, combined with the local gas density and water abundance, 
provide favorable conditions for maser action. In contrast to NM95, our
model naturally accounts for the clustering of the high-velocity maser spots 
into distinct clumps, the prominent dip in the maser spectrum at the systemic
velocity of the disk, and the striking asymmetry in the high-velocity maser 
spectrum of NGC 4258 in particular, and that of other masing disks in general.
In addition to its explanatory power, this model also has
some predictive power (\S{6.3}).
 
We should note that the Doppler-shift asymmetry in the spectrum of NGC 4258
may not pose a problem for the NM95 model, as Herrnstein \etal~(1996) 
pointed out that free-free absorption in the upper atmosphere of the disk may 
attenuate predominantly the blue-shifted maser beams due to the presumed
geometry of the warp (\S{1}). Such effect relies on a 
coincidentally special orientation of the warp with respect to the 
line-of-sight; it is unlikely to account for the Doppler-shift asymmetry 
for circumnuclear disks as a class, and in any case, it is not in conflict
with the model presented in the present investigation.

Finally, the model of NM95 yields a very low mass accretion rate of 
$7\!\times\!10^{-5} \alpha \solarmass \yr^{-1}$, where $\alpha\!<\!1$ is 
the dimensionless viscosity parameter introduced by Shakura and Sunyaev (1973).
Assuming a standard ``alpha disk'', where 
$\dot{M}\!\simeq\!3\pi\nu\Sigma$, and $\nu\!=\!\alpha h^2 \omega$ is the 
kinematic viscosity, we obtain for our model 
$ \dot{M} \!\simeq \! 7\!\times\!10^{-3} ~\! (\alpha/0.1) ~\! h_{16}^3 ~\!
n_9(\Hmol) \solarmass \yr^{-1}$. 
However, it is possible to estimate directly 
the mass accretion rate due to energy dissipation in the shocks: the energy
dissipation rate per unit length of the shock front is 
$\dot{E} dl \!=\! \Sigma ~\! v_s^3 ~\!dl/2$; the total length of the shock 
front within an annulus of width $dr$ is $m~\!dr/sin\pitch$, assuming 
the presence of $m$ ``grand-design'' spirals with pitch angle $\pitch$;
the kinematic viscosity is  $\nu\!=\! (2/3) r v_r$, where $v_r\!\approx\!r/t_r$
is the radial velocity of the gas, and  
$t_r\!\approx\!E/\dot{E}$ is the radial drift time-scale. We obtain 
\be \dot{M} ~\sim~ 7\!\times\!10^{-3}~ m~\! 
\left({v_s \over 20 \km\persec}\right)^2  n_9(\Hmol) ~\! h_{16}
\units \solarmass \year^{-1}. \label{mdot} \ee
We emphasize that equation (\ref{mdot}) gives only a rough 
estimate for the mass accretion rate due to energy dissipation in the
shocks since the spiral pattern cannot be uniquely determined (see \S{6.1.1}).
Yet, this result may suggest an advection-dominated accretion flow in NGC 4258,
as proposed by Lasota \etal (1996) to explain the observed spectrum from the
galaxy nucleus. They derived a mass accretion rate of 
$\dot{M}\!\sim\! 1.3\!\times\!10^{-3}~\!(\alpha/0.1) \solarmass \year^{-1}$, 
which is consistent with  equation
(\ref{mdot}) if $\alpha$ in their model is of 
order $0.5$.  We note, though, that Lasota \etal derived the {\it current\/}
mass accretion rate onto the black hole, while the above derivation
gives $\dot{M}$ at a radius of about $0.2\pc$ 
($\sim\!10^5$ Schwarzschild radii),
which need not necessarily be the same (it is possible that 
nucleus of NGC~4258 is relatively quiescent at the present time).

%%%%%%%%%%%%%%%%%%%%%%%%%%%%%%%%%%%%%%%%%%%%%%%%%%%%%%%%%%%%%%%
\section{OTHER CIRCUMNUCLEAR DISKS}
%%%%%%%%%%%%%%%%%%%%%%%%%%%%%%%%%%%%%%%%%%%%%%%%%%%%%%%%%%%%%%%

We have proposed that high-velocity water maser emission in NGC 4258 arises 
behind shock fronts in a largely non-inverted disk, and suggested that such
model may also account for the Doppler-shift asymmetry in other 
circumnuclear disks. We should emphasize that
water maser emission may also arise in circumnuclear disks 
under different circumstances, such as 
within some range of radii in a disk where the gas is largely inverted
(not enough IR photon trapping), or preferentially behind shock fronts in a 
largely inverted disk that maintains spiral activity.  
The prevalence of each situation is unknown, as it depends on 
the combined probability distribution of many characteristics of circumnuclear
disks and of their environment, as well as on biases in the selection of 
systems to be searched for high-velocity maser features. 
We argue that if the Doppler-shift asymmetry 
continues to be common in an increasingly larger number of identified 
circumnuclear disks, then the scenario proposed in 
the present paper is the most prevalent circumstance for the generation 
of detectable high-velocity maser emission in circumnuclear disks.
We now discuss some implications of our model to circumnuclear disks in 
general, and elaborate on the case of NGC 1068 in particular. 

High-velocity maser emission need not 
necessarily arise in every circumnuclear disk that exhibits low-velocity
emission, even if the disk is observed nearly edge-on. This may happen, among
other reasons, simply because there are no spiral shocks in 
the disk. The low-velocity features may arise in a masing annulus at the
inner edge of the disk which is directly irradiated by X-rays from 
a central source, as we suggest happens in NGC 4258 (\S{6.2}).
There is already one case - NGC 2639 - where a disk-origin of the 
low-velocity maser emission is indicated by their velocity drift
(Wilson, Braatz \and Henkel 1995), possibly caused due to centripetal 
acceleration as in NGC~4258, but a search for ``satellite,'' high-velocity 
maser features did not discover any.
 
High-velocity maser emission that originates behind spiral shock fronts 
need not necessarily exhibit the Doppler-shift asymmetry: the optical depth 
for absorption of the blue-shifted beams may be too small for producing 
noticeable asymmetry in the maser spectrum; 
the blue-shifted beams may not pass through the velocity coherent regions 
on the other side of the midline, where absorption occurs, if the combination
of disk thickness, inclination, and pitch angle of the spirals satisfy the
approximate condition $h/r<2\tan\pitch\cot\theta_{inc}$ 
(see Eq.~[\ref{define-integrand}]);
the disk may be largely inverted if the gas density and/or disk thickness 
are sufficiently small, in which case there would be no absorption.
As mentioned above, it is unclear how common these possible situations are. 

In cases where the high-velocity maser spectrum from a circumnuclear disk 
exhibits the Doppler-shift asymmetry, we predict that there should generally
be substantial absorption of maser emission near the systemic velocity of 
the disk.  We argued that it may account for the dip in the low-velocity
maser complex in NGC 4258 (\S{6.2}), and we suggest that it may also explain 
why the systemic maser emission in NGC 1068 and NGC 4945 is 
weaker than the red-shifted emission, in spite of probably being boosted 
due to illumination by a central background continuum source. 

According to our model, the energy that ultimately powers the high-velocity
masers is provided by energy dissipation in shocks, while the low-velocity 
masers probably arise near the inner edge of the disk which is irradiated 
directly by a central source. Since epochs of spiral shock activity in
a disk may not necessarily overlap with epochs of substantial nuclear 
activity, we predict that high-velocity maser emission could, in principle,
appear at the  absence of detectable low-velocity emission.  
Such situation may be detected in galactic nuclei that do not contain, 
at the present epoch, an
apparent central source that heats up the disk and which produces background 
radio continuum.  Obviously, it requires the existence of a molecular,
circumnuclear disk that is viewed nearly edge-on,
and which contains spiral shocks. 

We predict that high-velocity maser spots in circumnuclear disks that
exhibit the Doppler-shift asymmetry will appear clustered in 
clumps rather than being uniformly or randomly distributed, since
they arise where the lines-of-sight are tangent to the spiral shocks
(turbulent motions in the gas presumably cause the  
clumps to break up into individual spots).  
The clumps may be either narrow or broad, depending on whether the shock
fronts are smooth or heavily corrugated. The number of high-velocity maser
clumps depends on the spiral pattern, and could be as low as one.

%%%%%%%%%%%%%%%%%%%%%%%%%%%%%%%%%%%%%%%%%%%%%%%%%%%%%%%%%%%%%%%
%%%%%%%%%%%%   \subsection{NGC 1068}
%%%%%%%%%%%%%%%%%%%%%%%%%%%%%%%%%%%%%%%%%%%%%%%%%%%%%%%%%%%%%%%

The second most compelling case of a rotating structure, delineated by water
maser emission, is found in the nucleus of NGC 1068
(Gallimore \etal 1996; Greenhill \etal 1996).
The maser emission extends about $\pm\!300\km\persec$ from the galactic 
systemic velocity, tracing sub-Keplerian differential 
rotation around a central mass of $\sim\!10^7 \solarmass$ that lies within a
radius of about $0.7\pc$. The Doppler-shift
asymmetry is as striking as in NGC 4258, except that in NGC 1068 the 
red-shifted, high-velocity maser emission is stronger even than 
the low-velocity emission complex at the systemic velocity of the galaxy. 
The VLBI observations by Greenhill \etal (1996) currently provide the most
precise image of the disk, but only of its receding side.
It is delineated by five dominant clumps of maser emission, 
distributed in a nearly planar configuration on the sky: one at the 
systemic velocity of the galaxy, and the others at relatively 
red-shifted velocities. The weaker, blue-shifted, high-velocity features have
been recently imaged by Greenhill \etal (private communication), but the 
results are not yet available. 

The disk in NGC 1068 is not nearly as well-defined as the that in NGC 4258, 
and in fact, at present, it is not yet clear whether the observed structure 
is a disk or a thick torus.
Greenhill \etal (1996) proposed that the observed emission
traces part of the limb of an edge-on rotating torus with an opening angle
of about $90\deg$, rather than the
midplane of a disk as suggested by Gallimore \etal (1996).  
The reason for that is twofold: the radio jet axis is misaligned from the
nearly planar distribution of the red-shifted, high-velocity sources by
about $45\deg$ rather than $\sim\!90\deg$. 
In addition, the location of the blue-shifted emission, 
observed with lower resolution by Gallimore \etal (1996), is more consistent 
with being symmetric about the jet axis with respect to the red-shifted 
emission, then with all the features being nearly aligned in a plane. 

Yet, there are some difficulties with such model, as pointed out by Greenhill 
\etal (1996). The condition of hydrostatic equilibrium implies gas motions 
of order $100\km\persec$ which, if thermal, correspond to temperatures where
molecular gas could not survive. If the maser emission arises in cold gas
clouds with highly supersonic velocities, it is unclear why the high-velocity 
maser spots appear clustered in a few distinct ``super-clumps.'' 
It is also difficult to understand the absence of high-velocity maser emission 
from the southern half of the torus and near its midplane. 
Greenhill \etal suggested
that it may be related to the absence of strong southern jet emission.  But
if the irradiation of the inner surface of the torus is dominated by the jet, 
which is likely to produce also radio continuum emission, 
it is unclear why we do not detect systematic emission superimposed along some 
length of the jet, or at least along its northern part.
These problems do not arise if the rotating structure is a relatively thin 
disk, which cannot be ruled out at present since the uncertainties in the 
position of the weak, blue-shifted masers are quite large 
(Gallimore \etal 1996).  The axis of the radio jet may not be perpendicular 
to the masing disk, and its direction may be determined by the accretion 
disk on smaller scales, which could have a different orientation, 
or by the equatorial plane of a misaligned, spinning black hole.
VLBI observations of the blue-shifted emission would probably allow to 
distinguish between the torus and disk hypotheses.

We suggest that, as in NGC 4258, the high-velocity maser emission in 
NGC~1068 arises due to spiral shocks in a largely non-masing disk, which
accounts for the similar Doppler-shift asymmetry in the maser spectrum. 
The low-velocity masers may arise within a thin layer at the inner edge
of the rotating structure which is irradiated directly by the central source, 
and amplify a central background continuum source. 
Assuming the disk is relatively thin, 
the shock-origin scenario described in the present paper can account for 
the clustering of the high-velocity maser spots in NGC~1068 into a four 
clumps rather than being randomly distributed. However, unlike in NGC 4258, 
the shocks in NGC 1068 are probably heavily corrugated, as perhaps indicated by 
the larger extent of the clumps relative to the mean inter-clump separation.
The rotational speed in NGC~1068 is about three times lower than that in 
NGC~4258, so assuming the shock speed is at least as high as in NGC~4258, it is 
reasonable to expect the typical pitch angle of the spirals in NGC~1068 
to be $\gtrsim\!6\deg$ (Eqs.~[\ref{vs},\ref{shock-speed}]). 
Higher mean pitch angles allows larger possible variations in the 
local pitch angle throughout the spiral pattern. Thus, the higher the 
pitch angle, the more unlikely is that the maser clumps would all lie 
exactly along a single diameter through the disk. This may contribute to 
the observed deviations from a Keplerian rotation curve in NGC 1068, 
in addition to the probable effect of the star distribution on the 
gravitational potential (the volume enclosed within the radius of the 
masing disk in NGC~1068 is a few hundred times larger than the corresponding
volume in NGC 4258). 

The origin of the maser emission in NGC 1068, and the
role of absorption in producing the Doppler-shift asymmetry in this
system will be examined in more detail when VLBI images of the receding side of 
the disk become available, and the nature of the rotating structure
is revealed. 

%%%%%%%%%%%%%%%%%%%%%%%%%%%%
\section{SUMMARY}
%%%%%%%%%%%%%%%%%%%%%%%%%%%%
We propose an alternative paradigm for the origin of water maser emission 
in circumnuclear disks, where the high-velocity maser features arise behind
spiral shock fronts in a largely non-inverted disk. We summarize below the
main points, assumptions, results, and predictions of this investigation.

In the present investigation we assume molecular hydrogen 
densities of $\sim\!10^9 \cm^{-3}$, and show that population inversion 
may largely be quenched in circumnuclear disks due to trapping of the
IR photons produced in the pump cycle (\S{2}).
By contrast, Neufeld and Maloney (1995) suggested that the high-velocity 
maser emission in NGC 4258 originates in a largely inverted disk, where 
the molecular density is $\sim\!3\!\times\!10^7 \cm^{-3}$ (\S6.4). 

We suggest that spiral shocks may form in circumnuclear disks for the same 
reasons they do in spiral galaxies and in accretion disks, which is basically 
due to the presence of non-axisymmetric disturbances, of which there is no
shortage at the bottom of a galactic potential well (\S{3.2}). 
The post-shock regions provide ideal conditions for the generation of luminous
water maser emission: the energy that ultimately powers the maser
is naturally 
provided by the dissipation of the relative kinetic energies of the
shocked and unshocked gas; the post-shock chemistry produces copious 
amounts of water; the sheet-like geometry of shock fronts allows long
gain paths for amplification in the shock plane, and most important,
the IR photons can easily escape from the post-shock masing slab through the
shock front due to the velocity difference between the post-shock and
pre-shock gas, thus avoiding quenching the maser regardless of the 
dimension of the surrounding gas distribution (\S{3.1}). 

The high-velocity features arise in the post-shock masing slabs 
where the line-of-sight is tangent to the spiral fronts. 
The high velocities of the maser features reflect the circular 
motions in the disk, and have nothing to do with the shock speed.  
An immediate consequence of this scenario is that, regardless of the
azimuthal direction to the observer, the disk should appear to be 
delineated by clumps of high-velocity maser spots, as indeed observed, 
rather than by uniformly or randomly distributed spots (\S{4.1}).
The horizontal extent of the high-velocity clumps may be either narrow or
broad relative to the inter-clump separation, depending on whether the slabs
are heavily corrugated and thereby effectively thickened (\S{4.3}). 

The maser radiation travels through some part of the largely
non-masing disk along its way to the observer. Since the interaction of maser
photons with water molecules is a line process, the non-inverted gas
is transparent to the maser radiation, except in regions in the disk that
maintain close velocity coherence with the maser source, where
maser photons are absorbed. 
Since the spiral shocks have a trailing geometry due to the action of 
differential rotation, the high-velocity masers are always located in
front of the midline in the receding side of the disk,
and behind the midline in the
approaching side of the disk. The Doppler-shift asymmetry
arises because the blue-shifted maser photons are subject to absorption as
they pass through the velocity-coherent region on the other side of the
midline, while red-shifted photons never cross such region since the l.o.s
velocity in the disk declines monotonically along their path to the observer
(\S{5.1}). This is effect is suggested to account
for the observed Doppler-shifted asymmetry in the high-velocity maser spectra
from circumnuclear disks.

The expected isotropic luminosities of the high-velocity maser 
features is calculated assuming either C-type or dissociative 
J-type shocks (\S{4.2}). The attenuation of the blue-shifted,
high-velocity beams due to either saturated or unsaturated absorption is 
derived in \S5.2.
We find that the observed maser luminosities of the blue-shifted and 
red-shifted, high-velocity features in NGC~4258 are consistent with the
predicted from our model, the more so in the case of dissociative,
J-type shocks (\S6.1). 
The shock type cannot be determined {\it a priori\/} since
it depends upon such poorly known parameters as the preshock magnetic 
field and the length scale of the ion-neutral coupling.
 
The typical pitch angle of the spirals in NGC~4258 is probably about
two degrees, and the typical shock speed is about $20\km\persec$ (\S{6.1.2}).
The global pattern of the spiral shocks cannot be uniquely determined, but
we note that a ``grand-design'' pattern which consists of two, roughly
equally-spaced spirals can reproduce the observed regularity in the spacing
between the high-velocity clumps (\S{6.1.1}). The high-velocity features
arise nearly along a diameter through the disk that makes an angle of about 
two degrees with the midline, 
which does not introduce a noticeable deviation from a Keplerian
rotation curve. The velocity gradient associated with the shock itself is
always perpendicular to the l.o.s at a maser location.
The corrections implied by this model to the previously inferred black hole mass
and galaxy distance are negligible (\S{6.1.1}).

The proposed scenario for the generation of the Doppler-shift asymmetry
is based on a few basic assumptions: (1) The molecular density
in the disk should not exceed a few times $\sim\!10^9 \cm^{-3}$, depending 
exactly on the shock type and strength, in order  not to collisionally
quench inversion in the post-shock gas (\S{3.1}), and 
the density should also not be much lower
than a few times $10^8 \cm^{-3}$, depending exactly on the dimensions of the
disk and the water abundance, in order for the gas in the disk to be largely 
quenched due to IR photon trapping (\S{2}). In lower density regions in the
disk, where the level populations may be largely inverted, maser emission may
still arise preferentially in the warm post-shock gas, but
in these regions we do not expect any systematic difference between the 
brightness of the red-shifted and blue-shifted, high-velocity features.
(2) The
shock speed needs to be high enough above the sound speed and the turbulent
velocity in order to allow efficient escape of IR photons from the warm
post-shock region through the shock front (\S{4.1}). (3)
The pitch angle of the spiral should not be too high, depending exactly on the
disk thickness and inclination, in order for the blue-shifted maser beams to
pass through the corresponding velocity-coherent regions on the other side of
the midline, where absorption occurs (\S\S5.1, 7). Finally, we assume that (4)
the gas and dust are nearly in thermal equilibrium, so re-processing of the IR
photons (produced in the pump cycle) by the dust does not significantly 
affect the level populations of the water maser transition
as to become inverted. 

According to our model, the mass of the masing disk in NGC~4258 is 
$\sim\!10^4 \solarmass$, and the
mass accretion rate due to energy dissipation in the shocks is of order
$\sim\!7\!\times\!10^{-3}\solarmass \year^{-1}$, which is
consistent with the prediction of the advection-dominated accretion flow model 
of Lasota \etal (1996).
By contrast, the disk mass in the model of Neufeld and Maloney (1995) is
$\sim\!100 \solarmass$, and the  mass accretion rate is 
$\dot{M}\!\sim\!7\!\times\!10^{-5}\alpha$ (\S{6.4}).

We suggest that the low-velocity emission arises within a narrow annulus 
near the inner edge of the largely non-masing disk, where direct irradiation by 
a central source may provide the energy which ultimately powers the maser.
This explains why the observed low-velocity maser sources in NGC~4258 
are confined to a narrow annulus at the inner edge of the disk, even-though 
velocity coherence is maintained along the entire length of the minor axis 
of the projected disk. It may also account for the prominent dip in the maser 
spectrum of NGC~4258 near the systemic velocity of the disk, as being due to 
substantial absorption along the line-of-sights to the center of the disk.
The absence of high-velocity maser features at the inner edge of the disk
($0.13\pc$) is because the velocity coherent regions within the masing
annulus are likely to be elongated in the radial direction, and due to the 
amplification of the central background continuum source (\S{6.2}).

The shock-origin scenario  has a unique observational prediction:
since the high-velocity masers arise where the \los is tangent to 
spirals, the slow rotation of the waves must result in a continuous,
outward drift in the positions of all the high-velocity maser features, as
well as 
in a slow decrease in their \los velocities. Since these systematic effects 
must be shared by all the high-velocity maser features, it may be 
possible to test this prediction by a careful statistical analysis
in NGC~4258, where the rotational 
velocities, and thus the drift rates, are the highest known. 

We discuss implications of the present model to other circumnuclear disks
in \S7,  and note that water maser emission may also arise in 
circumnuclear disks under different circumstances. High-velocity 
maser emission need not necessarily arise in every circumnuclear
disk that exhibits low-velocity emission, and in principle, 
high-velocity maser could appear at the absence
of detectable low-velocity emission.  
The prevalence of each situation is unknown, as it depends on
the combined probability distribution of many disk characteristics and 
environmental conditions. 
We predict, though, that in cases where the high-velocity maser 
spectrum of a circumnuclear disk
exhibits the Doppler-shift asymmetry, there should generally
be substantial absorption of maser emission near the systemic velocity of
the disk, and that
the high-velocity maser spots would appear clumped rather than
randomly distributed.  
Assuming that the Doppler-shift asymmetry observed in the high-velocity
spectrum of the masing disk in NGC~1068 is also due to attenuation of 
the blue-shifted beams, we predict that the rotation structure 
on a sub-parsec scale in NGC~1068 is a relatively thin disk rather than a torus.
The shock-origin scenario in this system
will be examined in more detail when VLBI images of the blue-shifted
side of the disk become available, and the nature of the rotating structure
is revealed.

Finally, we note that extragalactic high-velocity maser emission need 
not necessarily always arise in a rotating disk (\S{1}).
It may arise in outflows from galactic nuclei, in which case
the blue-shifted features are likely to appear stronger because the
radially-approaching side of the flow may amplify a background continuum
source at the center of expansion, as found in a case of interstellar
masers (Gwinn 1994).  It may arise along
radio jets, possibly at shock interfaces between the jet and the dense
near-nucleus gas, as found along the jet of NGC 1068 (Gallimore \etal 1996).
Furthermore, if the maser emission originates in the vicinity of a massive 
black hole, whether in a disk, an outflow, or jets, then identifying features 
as ``systemic'' or ``satellite'' can be confused by the possible motion of
the black hole 
within the galaxy nucleus, which would shift the maser spectrum 
in either direction by up to the velocity dispersion in the galaxy nucleus.
Thus, evidence for a disk origin of maser emission must rely on
both the kinematic and spatial distribution of the maser sources.

\vspace{0.2truein}
\acknowledgements
Special thanks to David Neufeld for providing us with the energy levels of 
water, and his unpublished results for the maser rate coefficient $Q$ at very
high $\xi$ values, as well as for useful discussions. 
We also thank Moshe Elitzur for comments and discussions,
and David Hollenbach for discussions. 
The research of CFM is supported in part by National Science Foundation grant,
AST95-30480.

%%%%%%%%%%%%%%%%%%%%%%%%%%%%%%%%%%%%%%%%%%%%%%%%%%%%%%%%%%%%%%%
\newpage

\begin{figure}[ht]
\vskip-0.7truein
\textwidth=3in
 \centerline{
\psfig{figure=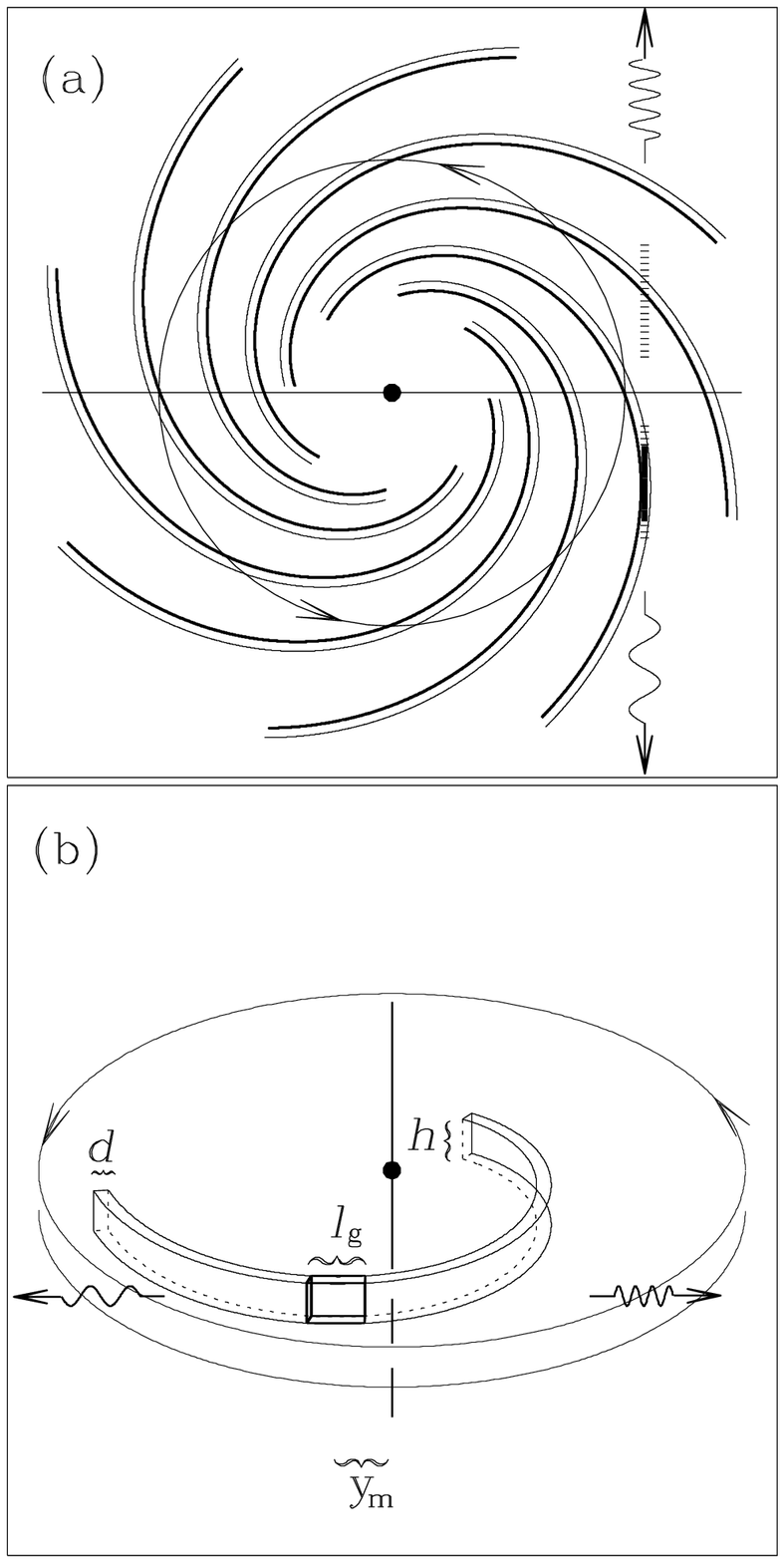,height=7.0in,width=7.0in}}
\caption{ 
Illustration of the general geometry: (a) 
the spiral regions represent the thin masing slabs behind shock fronts, the
horizontal line is the midline - the diameter through the disk perpendicular
to the l.o.s, and the dark rectangular, located where the \los is tangent to a
spiral represents a high-velocity maser (\S{4.1}). 
Since the disk is largely not-masing, 
photons from that maser are absorbed in the shaded regions, 
where close velocity coherence is maintained 
with the maser source. The Doppler-shift asymmetry arises due to
the trailing geometry of the spirals, which gives rise
to excess absorption of the blue-shifted radiation (\S{5.1}). 
(b) the high-velocity maser emission originates in a box-shaped region of
thickness $d\!\sim\!10^{13-14}/n_{H,9} \cm$, height $h$, and 
length $l_g$ which is determined
by $d$ and the local radius of curvature of the spiral. It is located at 
a distance $|y_m|\!\simeq \!r\sin\pitch$ from the midline, where $\pitch$ is 
the pitch angle of the spiral (\S\S{4.3.1}, {4.1}).
}
\end{figure}

\begin{figure}[ht] 
\vskip-0.7truein 
\textwidth=3in 
 \centerline{ 
\psfig{figure=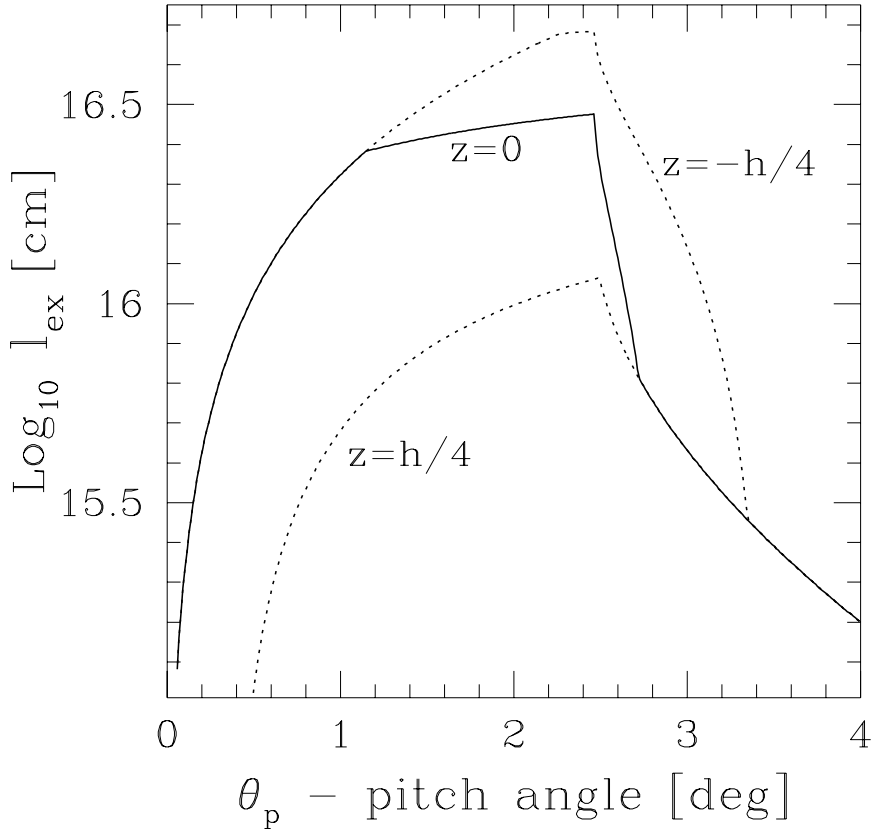,height=7.0in,width=7.0in}} 
\caption{ 
The velocity coherent path-length for excess absorption of blue-shifted,
high-velocity radiation, $\excess(z)$, at the mean radius of the disk in
NGC 4258, as a function of pitch angle (Eq.~[\ref{define-excess}]). 
It assumes a disk thickness of $10^{16}\cm$, 
and maser line-width of $2\km\persec~$ (see text). The three curves 
correspond to maser radiation emanating from the maser at the midplane 
of the disk ($z\!=\!0$), and at half way up (and down) in the disk atmosphere 
($z\!=\!\pm h/4$). The sharp
decline at $\pitch\!\gtrsim\!3\deg$ occurs because the line-of-sight,
which is inclined to the disk plane by $7\deg$, 
does not pass through the absorbing region on the other side of the midline. 
The estimated pitch angle in NGC 4258 is of order two degrees (\S{6.1.2}).
}
\end{figure} 

\begin{figure}[ht] 
\vskip-0.7truein 
\textwidth=3in 
 \centerline{ 
\psfig{figure=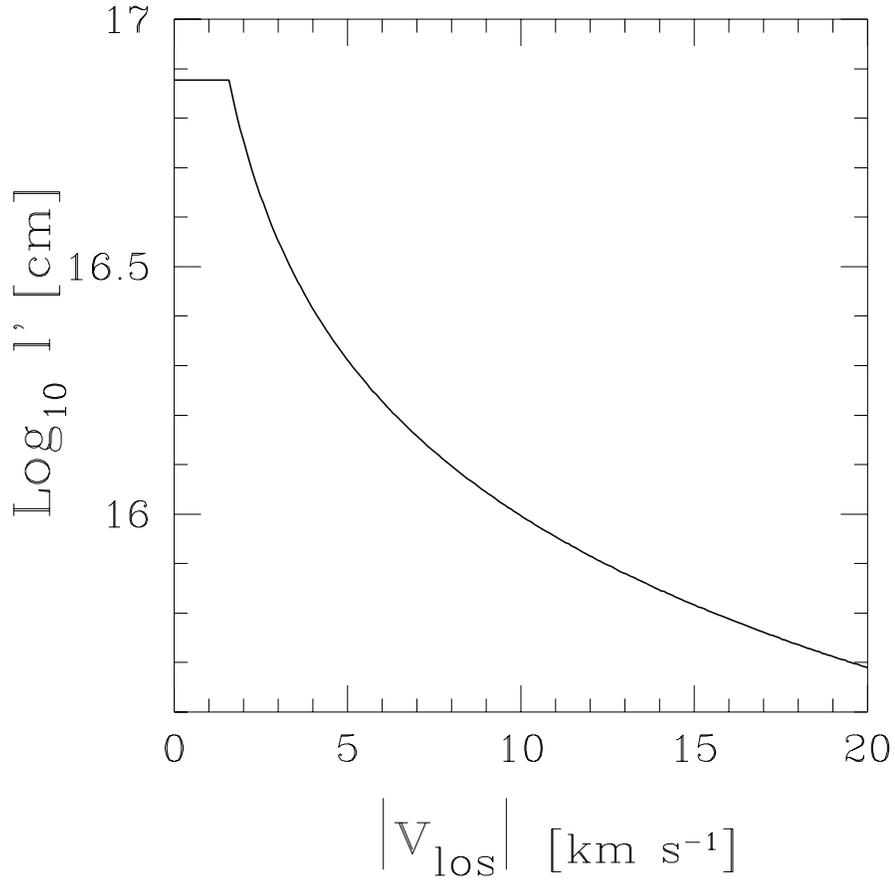,height=7.0in,width=7.0in}} 
\caption{ 
The length of the absorbing column of gas in the foreground
of the low-velocity masers in NGC 4258, as a function of $\vlos$, 
assuming maser linewidth of $2\km\persec$, disk thickness of $0.003\pc$, and 
inclination $\theta_{inc}\!\!=\!83\deg$ (Eq.~[\ref{lowv-zeta}]).
The absorption of low-velocity maser emission, which drops rapidly with 
increasing relative velocity, may explain the prominent 
dip in the maser spectrum at the systemic velocity of the disk (\S{6.2}).
}
\end{figure}

%%%%%%%%%%%%%%%%%%%%%%%%%%%%%%%%%%%%%%%%%%%%%%%%%%%%%%%%%
%             R E F E R E N C E S                       %
%%%%%%%%%%%%%%%%%%%%%%%%%%%%%%%%%%%%%%%%%%%%%%%%%%%%%%%%%
\newpage
\def\MNRAS{MNRAS}

%%%%%%%%%%%%%%%%%%%%%%%%%%%%%%%%%%%%%%%%%%%%%%%%%%%%%%%%%%%%%%%%%%%%%%%%
\end{document}